\documentclass{article}

\usepackage[breaklinks]{hyperref}
\usepackage{lineno}
\modulolinenumbers[5]

\usepackage[utf8]{inputenc}
\usepackage{amsmath}
\usepackage{psfrag}
\usepackage{subfig}
\usepackage{color}
\usepackage{empheq}
\usepackage{tikz}
\usepackage{verbatim}
\usepackage{textcomp}

\newcommand{\bs}{\boldsymbol}

\begin{document}

\title{Continuum Mechanical Modeling of Strain-Induced Crystallization in Polymers}

\author{\\ Serhat Aygün\thanks{Corresponding author, email address: serhat.ayguen@tu-dortmund.de} , Sandra Klinge \\ \\
{\normalsize Institute of Mechanics, TU Dortmund University,} \\
{\normalsize Leonhard-Euler-Strasse 5, 44227 Dortmund, Germany} \\[1ex]}

\date{ }

\maketitle

\begin{abstract}
The present contribution focuses on the thermodynamically consistent mechanical modeling of the strain-induced crystallization in unfilled polymers. This phenomenon is of particular importance for the mechanical properties of polymers as well as for their manufacturing and the application. The model developed uses the principle of the minimum of dissipation potential and assumes two internal variables: the deformations due to crystallization and the regularity of the network. In addition to the dissipation potential necessary for the derivation of evolution equations, the well-established Arruda-Boyce model is chosen to depict the elastic behavior of the polymer. Two special features of the model are the evolution direction depending on the stress state and the distinction of crystallization during the loading and unloading phase. The model has been implemented into the finite element method and applied for numerical simulation of the growth and shrinkage of the crystal regions during a cyclic tension test for samples with different initial configurations. The concept enables the visualization of the microstructure evolution, yielding information that is still inaccessible by experimental techniques.
\end{abstract}
~ 

\newpage

\section{Introduction}

\begin{figure}[htb]
  \psfrag{lam}[]{\footnotesize Stretch $\lambda$}
  \psfrag{stress}[][][1][180]{\footnotesize Stress [MPa]}
  \psfrag{cd}[][][1][180]{\scriptsize \hspace{2mm} Crystallinity degree [\%]}
 \centering
 (a)\includegraphics[width=0.44\textwidth]{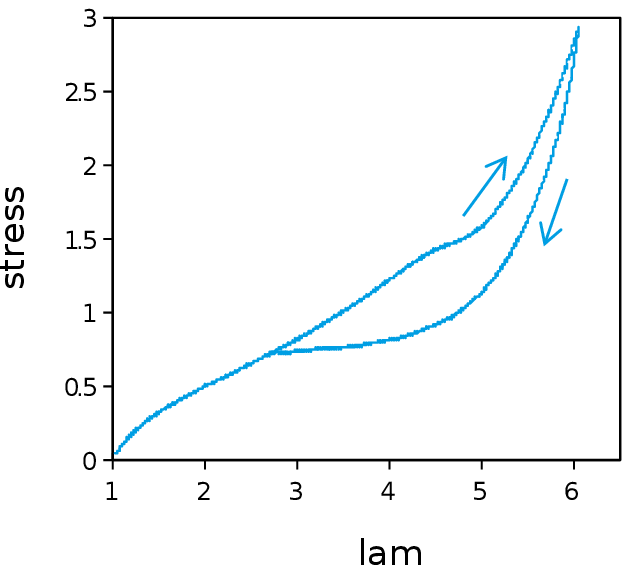} \hspace{1mm}
 (b)\includegraphics[width=0.44\textwidth]{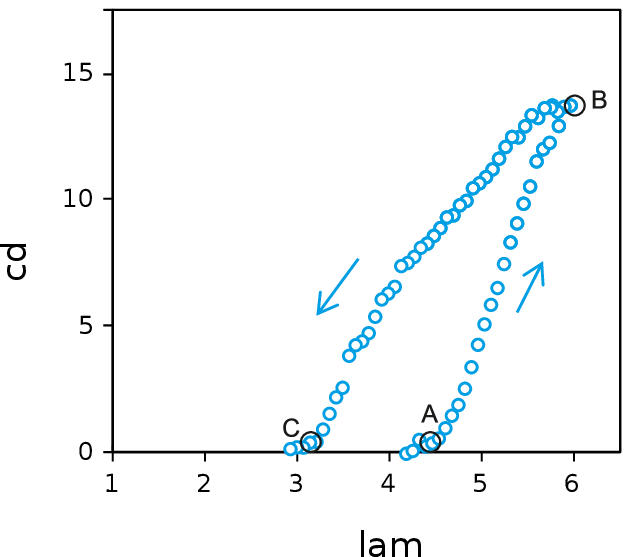}
  \caption {Uniaxial cyclic tensile test for unfilled  natural rubber at room temperature and at a constant strain rate of $4.2 \cdot 10^{-3}$ s$^{-1}$ \cite{CANDAU2015244}. (a) Stress-stretch diagram. (b) Crystallinity degree-stretch diagram.} 
  \label{fig1}
\end{figure}

The strain-induced crystallization (SIC) typically occurs in filled and unfilled natural and synthetic rubbers. The main characteristic of this phenomenon is that the high strains cause the development of crystalline regions within the original amorphous polymer matrix, which increases tensile strength and considerably improves  crack growth resistance \cite{mistry2014}.
SIC can be characterized by techniques such as volume change measurements \cite{doi:10.1063/1.1746537}, stress relaxation \cite{TF9545000521}, birefringence \cite{doi:10.5254/1.3540495}, infrared absorption \cite{Siesler:85}, pulsed NMR \cite{doi:10.1080/00222348108015313}, dilatometry \cite{doi:10.5254/1.3536126}, electron microscopy \cite{Andrews232} and the small-angle X-ray scattering (SAXS) \cite{Kojio2011}. An alternative method providing significant information on crystalline content, crystallite size and orientation is the in situ wide-angle X-ray diffraction (WAXD)  \cite{katz1925,MITCHELL19841562,doi:10.5254/rct.13.86977}. Among others, the method has been used  by Tosaka \cite{TOSAKA2012864} to study  the SIC kinetics and in the work by Brüning \cite{bruening2012} to investigate effects of dynamic load.
 Moreover, the work by Candau et al. (2014) \cite{doi:10.1021/ma5006843} aims to identify the domains of the material involved in the SIC process and to quantify the length of the chains in these areas by using WAXD. A methodology is also proposed to quantify the distribution of the local network densities and the distribution of the corresponding crystallite size. The influence of the crystallization history of the material on its recrystallization ability is also investigated. However, almost all of the mentioned techniques have been performed with sequential measurements, whereby the specimen is first expanded to a desired strain, fixed at this strain, and then the specimen is removed from the stretcher. In a last step, it is clamped on the analyser to perform the examination. The work by Toki et al. (2000) \cite{TOKI20005423} presents an especially developed instrument with which it is possible to continuously measure the stress-strain behaviour and the X-ray scattering intensity simultaneously during the expansion and the subsequent retraction. An alternative method to the conventional X-ray diffraction is proposed by Le Cam (2018) \cite{doi:10.1111/str.12256}. The new method is based on temperature measurement and quantitative calorimetry to determine the crystallinity of rubber in mechanical tests. For this purpose, the heat power density is first determined from temperature variation measurements and the heat diffusion equation. Finally, the crystallinity is calculated from the temperature variations caused by the SIC.
\\ \\
The representative results of a cyclic test performed for the unfilled  natural rubber under constant temperature and speed are shown in Fig. \ref{fig1}. Here, the stress diagram (Fig. \ref{fig1} a) builds a hysteresis thus indicating the dissipative nature of the SIC phenomenon. The volume fraction of crystalline regions, the so-called crystallinity degree, is also used to monitor the process of crystallization. According to Fig. \ref{fig1} b, the crystalline regions start to build after a threshold $\lambda_A$ is exceeded. Thereafter, the crystalline regions grow/nucleate and the crystallinity degree increases. The upper bound of stretches is $\lambda_B$. An increase of overall deformations and exceeding of maximum $\lambda_B$ yields the inelastic deformations, which is not the subject of this study. The reduction of the crystallinity degree during the unloading phase is less intensive than its growth by the loading: The material becomes completely amorphous at stretch $\lambda_C < \lambda_A$.
\\ \\
The first attempts to give a quantitative expression to the effect of deformations on crystallization were made by Alfrey and Mark (1942) \cite{alfrey1942}. Their treatment only applies to a single chain and therefore cannot be associated with the extended network structure. On the basis of the aforementioned work, Flory (1947, 1949) \cite{doi:10.1063/1.1746537,flory1949} formulated the classic thermodynamic theory of SIC. In his contributions, the degree of crystallization is expressed in terms of the change in the crystallization temperature which, in turn, depends on the elongation. Later on, a phenomenological expression to describe the growth of the crystalline phase was developed by Doufas et al. (1999) \cite{doufas1999} using a modified Avrami equation. The work by Ahzi et al. (2003) \cite{ahzi2003} closely follows the contribution by Boyce et al. (1993) \cite{ARRUDA1993389}, where the elastic resistance is modeled by a combined process of molecular relaxation and network orientation. In contrast to the work by Boyce, however, the focus is on strain-induced crystallization, where the previously mentioned phenomenological expression by Doufas et al. (1999) is used to explain the growth of the crystalline phase. The authors Negahban (2000) \cite{NEGAHBAN20002811} and Rao and Rajagopal (2001) \cite{RAO20011149} used large deformations in continuum mechanics to model the two phases, amorphous and crystalline, separately during a uniaxial stretching of the polymer. Afterwards, Tosaka et al. (2004) \cite{tosaka2004} developed a micromechanical model that takes the existence of different chain lengths into account. Accordingly, the shorter chains are completely stretched under tension and subsequently form nucleation sites for crystallites. An extended model was developed by Kroon (2010) \cite{kroon2010}, which also deals with the anisotropic nucleation of unfilled rubber. In this model, the amorphous phase is assumed to be viscoelastic. Under the assumption that the dissipative process is not only due to the crystallization but also to the viscoelastic behavior, the model is able to predict both the stress-strain hysteresis and the development of the degree of crystallization under deformation. A sophisticated micromechanical continuum model for partially crystallized polymers was developed by Mistry and Govindjee (2014) \cite{mistry2014}. Here, the use of micro-macro-transition theories is new. The micromechanical model is connected to the macroscopic level using the non-affine microsphere model. The model is able to quantitatively predict the macroscopic behavior of strain-crystallizing rubbers. Recently, in a similar way the authors Nateghi et al. (2018) \cite{nateghi2018} proposed a micromechanical model which is incorporated into the affine microsphere model. In a further contribution by Dargazany et al. (2014) \cite{dargazany2014}, an extended micromechanical model for SIC in filled rubbers was presented. Besides the stress build-up and the evolution of crystallinity, inelastic properties of filled rubbers such as the Mullins effect, the permanent setting effect and the induced anisotropy are considered. The model shows good agreement with the experimental results, both in terms of stress elongation and crystallization-stretch relationships. In a recent publication by Behnke et al. (2018) \cite{BEHNKE201815}, the time and temperature dependence of SIC is modeled taking into account the induced anisotropy.
\\ \\
As the previous overview shows, the development of measurement techniques has already made a significant contribution to the investigation of the SIC process. Nevertheless, there are still open issues which have not yet been sufficiently clarified by experimental studies. Among others, the following issues can be pointed out: the form and distribution of crystalline regions in the material for high-strain states; the influence of the crosslinking degree of polymer chains and the interaction of crystalline regions. These phenomena are related to the nanoscale and are thus still not accessible by the experimental techniques. Moreover, already existing mechanical models mostly provide data on the effective material behavior without giving insight into the developed microstructure. Different from these strategies, the present model treats the microstructural changes in more detail and enables the simulation of amorphous polymer as well as of crystalline regions. The main goal of such an approach is to visualize the development of a microstructure within a representative volume element and to investigate its dependence on the external influences. The approach  primarily focuses on the simulation of unfilled polymers, the properties of which are presented in Fig. \ref{fig1}. These polymers are nearly incompressible materials in contrast to the filled  polymers where the volume change is slightly larger. The straightforward extension of the model capturing the material compressibility is also discussed.
\\ \\
The present contribution is structured as follows. Section \ref{sec1} introduces internal variables to simulate the SIC and deals with the thermodynamic consistency of the Helmholtz free energy density. The latter includes the Arruda-Boyce model as a basis. This approach applies to finite deformations and relies on a split into a volumetric and a deviatoric part. The study of the thermodynamic consistency starts with the Clausius-Duhem inequality and eventually shows definitions of conjugate pairs. Section \ref{sec2} focuses on the derivation of evolution equations for the internal variables. For this purpose, the principle of minimum of the dissipation potential is applied. This generic procedure is complemented by two assumptions: The first one couples the internal variables, whereas the second one concretizes the form of the dissipation potential. Both are chosen such that the resulting evolution equation simulates the increase and the decrease of the crystalline regions during loading and unloading. Furthermore,  Sect. \ref{sec3} discusses numerical aspects and the algorithmic treatment for the implementation of the SIC material model. In addition, Appendix A provides details on the FEM-implementation in the case of material and geometrical nonlinearity. Finally, selected numerical examples pertaining cyclic tensile loads visualize the microstructure evolution (Sect. \ref{sec4}). The first two academic examples are performed to study the influence of separate factors on the growth of crystalline regions. Thereafter, two case studies monitor the microstructure evolution for realistically chosen samples. These tests finally provide results for stresses and crystallinity degree which are validated according to the experimental data. The paper finishes with conclusions and an outlook.

\section{Assumption for the free energy density and check of thermodynamic consistency}
\label{sec1}

\subsection{Definition of internal variables}
The process of SIC occurs in the case of large deformations, such that the application of the theory of finite deformations is necessary for its reliable simulation. Typical of this theory, the modeling of dissipative processes is achieved by a multiplicative decomposition of the deformation gradient $\bs{F}$, which in our case incorporates an elastic part ($\bs{F}^{\mathrm{e}}$) and a part due to the crystallization ($\bs{F}^{\mathrm{c}}$)
\begin{linenomath*}
\begin{equation}
  \bs{F} = \bs{F}^{\mathrm{e}} \cdot \bs{F}^{\mathrm{c}} \text{ .}
  \label{FF}
\end{equation}
\end{linenomath*}
However, the description of the SIC process requires the introduction of an additional internal variable which, in the present model, determines the regularity of the polymer chain network, furthermore denoted by $\chi$. Regularity in this context implies the information about the orientation of the polymer chains to each other and the degree of order among the polymer atoms. Variable $\chi$ takes the value from the range $[0,1]$, such that values close to zero correspond to an amorphous state, whereas values close to the value of one are classified as crystalline regions. During the tensile test, the regularity evolves thus simulating the formation/degradation of crystalline regions.

\subsection{Assumption for the free energy density}
The free energy density assumed consists of two terms: an elastic part and a part due to crystallization
\begin{linenomath*}
\begin{equation}
  \Psi(J^{\mathrm{e}}, \bs{C}^{\mathrm{e}}, \chi) = \Psi^{\mathrm{e}} (J^{\mathrm{e}}, \bs{C}^{\mathrm{e}}) + \Psi^{\mathrm{c}}(\chi) \text{ .}
  \label{psi}
\end{equation}
\end{linenomath*}
The first term corresponds to the elastically stored energy and additively splits volumetric and deviatoric contributions
\begin{linenomath*}
\begin{equation}
\begin{split}
  &\Psi^{\mathrm{e}}(J^{\mathrm{e}}, \bs{C}^{\mathrm{e}}) = \Psi^{\mathrm{vol}}(J^{\mathrm{e}}) + \Psi^{\mathrm{dev}}(\bs{C}^{\mathrm{e}}) \text{ ,} \quad \Psi^{\mathrm{vol}}(J^{\mathrm{e}}) = K \, U(J^{\mathrm{e}}) \text{ ,} \\
  & U(J^{\mathrm{e}}) = \frac{1}{4} \left( (J^{\mathrm{e}})^2 - 1 - 2 \, \mathrm{ln}(J^{\mathrm{e}}) \right) \text{ ,} \quad J^{\mathrm{e}} = \mathrm{det}(\bs{F}^{\mathrm{e}}) \text{ ,} \quad \bs{C}^{\mathrm{e}} = {\bs{F}^{\mathrm{e}}}^T \cdot \bs{F}^{\mathrm{e}} \text{ .}
\end{split}
  \label{psi_split}
\end{equation}
\end{linenomath*}
Here, $K$ denotes the bulk modulus, $J^{\mathrm{e}}$ is a measure of the elastic volume change and $\bs{C}^{\mathrm{e}}$ is the elastic right Cauchy-Green tensor. Expression $K \, U(J^{\mathrm{e}})$ is the volumetric part of the energy. The term $\Psi^{\mathrm{dev}}$ corresponds to the Arruda-Boyce model \cite{ARRUDA1993389}
\begin{linenomath*}
\begin{equation}
\begin{split}
  &\Psi^{\mathrm{dev}} (\bs{C}^{\mathrm{e}}) = \mu \, \lambda_m^2 \left( \frac{\lambda_{\mathrm{chain}}}{\lambda_m} \beta + \mathrm{ln} \frac{\beta}{\mathrm{sin}(\beta)} \right) \text{ ,} \quad \beta = L^{-1}\left( \frac{\lambda_{\mathrm{chain}}}{\lambda_m}  \right) \text{ ,} \\
  &\lambda_{\mathrm{chain}} = \sqrt{\frac{\bar{I}_1}{3}} \text{ ,} \quad \bar{I}_1 = J^{e^{-\frac{2}{3}}} \mathrm{tr}(\bs{C}^{\mathrm{e}})  \text{ .}
\end{split}
  \label{ab_orig}
\end{equation}
\end{linenomath*}
Equation \eqref{ab_orig}
is  a constitutive relationship for the nonlinear elastic deformation of rubber materials  and  does not involve the effects of the SIC phenomenon. It is based on the eight-chain model capturing the influence of the rubber network microstructure. In the previous material law, $\mu$ denotes the shear modulus, $\lambda_m$ is the limiting network stretch, $\lambda_{\mathrm{chain}}$ is the chain stretch depending on the deviatoric first invariant $\bar{I}_1$ and $\beta$ denotes the inverse Langevin function which is related to the energy of a single random chain. The latter function cannot be expressed explicitly and is usually approximated by the Taylor series truncated up to the certain order  \cite{TF9545000881,doi:10.1177/1081286511429886}. The present work assumes that an approximation
including three terms of the Taylor series provides a sufficient accuracy. In this case, the Arruda-Boyce energy takes the form
\begin{linenomath*}
\begin{equation}
\begin{split}
  &\Psi^{\mathrm{dev}} (\bs{C}^{\mathrm{e}}) = \frac{\bar{\mu}}{2} \left[ \left( \bar{I}_1 - 3 \right) + \frac{1}{10 \, \lambda_m^2} \left( \bar{I}_1^2 - 9 \right) + \frac{11}{525 \, \lambda_m^4} \left( \bar{I}_1^3 - 27 \right) \right] \text{ ,} \\
  & \bar{\mu} = \frac{\mu}{1 + \frac{3}{5 \, \lambda_m^2} + \frac{99}{175 \, \lambda_m^4}} \text{ .}
\end{split}
  \label{ab}
\end{equation}
\end{linenomath*}
Alternatively to the Taylor series, a range of Pad{\'e} approximations can be applied  for the numerical evaluation of the inverse Langevin function. These approximations have different degrees of accuracy and complexity as discussed in the review papers by Jedynak  \cite{Jedynak} and Carroll  \cite{doi:10.1098/rsta.2018.0067}.
The second term in Eq. \eqref{psi} is assumed to depend linearly on the regularity
\begin{linenomath*}
\begin{equation}
\Psi^{\mathrm{c}}(\chi) = c \, \chi \text{ ,}
\end{equation}
\end{linenomath*}
and it has a crucial role in distinguishing the loading and unloading mode as shown in Sect. \ref{load_unload}.

\subsection{Thermodynamic consistency}
\label{thermo}
The Coleman–Noll procedure \cite{Coleman1963} is faced with the problem of finding necessary and sufficient conditions ensuring that the dissipation inequality is satisfied. The second law of thermodynamics for a purely mechanical process is expressed locally by the Clausius–Duhem inequality \cite{Cimmelli911}
\begin{linenomath*}
\begin{equation}
  D = \bs{P} \colon \dot{\bs{F}} - \dot{\Psi} = \bs{P} \colon \dot{\bs{F}} - \frac{\partial \Psi^{\mathrm{e}}}{\partial \bs{F}^{\mathrm{e}}} \colon \dot{\bs{F}^{\mathrm{e}}} - \frac{\partial \Psi^{\mathrm{c}}}{\partial \chi} \dot{\chi} \ge 0 \text{ ,}
  \label{di1}
 \end{equation}
\end{linenomath*}
where $D$ is the dissipation, $\bs{P}$ is the first Piola-Kirchhoff stress tensor and $\bs{P} \colon \dot{\bs{F}}$ is the internal power. The elastic deformation gradient is dependent on the deformation gradient and the gradient due to the crystallization $\bs{F}^{\mathrm{e}}(\bs{F},\bs{F}^{\mathrm{c}})$ (see Eq. \eqref{FF}), such that the rate $\dot{\bs{F}^{\mathrm{e}}}$ can be written as
\begin{linenomath*}
\begin{equation}
  \dot{\bs{F}^{\mathrm{e}}} = \dot{\bs{F}} \cdot {\bs{F}^{\mathrm{c}}}^{-1} + \bs{F} \cdot \dot{\bs{F}^{\mathrm{c}}}^{-1} = \dot{\bs{F}} \cdot {\bs{F}^{\mathrm{c}}}^{-1} - \bs{F}^{\mathrm{e}} \cdot \dot{\bs{F}^{\mathrm{c}}} \cdot {\bs{F}^{\mathrm{c}}}^{-1} \text{ .}
  \label{dFe}
\end{equation}
\end{linenomath*}
The inverse of $\dot{\bs{F}^{\mathrm{c}}}$ is determined by  taking the time derivative of the unity tensor $\bs{I}$ 
\begin{linenomath*}
\begin{equation}
  \bs{F}^{\mathrm{c}} \cdot {\bs{F}^{\mathrm{c}}}^{-1} = \bs{I} \quad \Rightarrow \quad \dot{\overline{\left(\bs{F}^{\mathrm{c}} \cdot {\bs{F}^{\mathrm{c}}}^{-1} \right)}} = \bs{0} \text{ ,}
\end{equation}
\end{linenomath*}
which, after rearranging, yields the expression
\begin{linenomath*}
\begin{equation}
\dot{\bs{F}^{\mathrm{c}}}^{-1} = - {\bs{F}^{\mathrm{c}}}^{-1} \cdot \dot{\bs{F}^{\mathrm{c}}} \cdot {\bs{F}^{\mathrm{c}}}^{-1} \text{ .}
\end{equation}
\end{linenomath*}
By inserting Eq. \eqref{dFe} into Eq. \eqref{di1}, the dissipation inequality becomes
\begin{linenomath*}
\begin{align}
  D = \bs{P} \colon \dot{\bs{F}} - \frac{\partial \Psi^{\mathrm{e}}}{\partial \bs{F}^{\mathrm{e}}} \colon \dot{\bs{F}} \cdot {\bs{F}^{\mathrm{c}}}^{-1} + \frac{\partial \Psi^{\mathrm{e}}}{\partial \bs{F}^{\mathrm{e}}} \colon \bs{F}^{\mathrm{e}} \cdot \dot{\bs{F}^{\mathrm{c}}} \cdot {\bs{F}^{\mathrm{c}}}^{-1} - \frac{\partial \Psi^{\mathrm{c}}}{\partial \chi} \dot{\chi} &\ge 0 \\
  \Rightarrow \, \left( \bs{P} - \frac{\partial \Psi^{\mathrm{e}}}{\partial \bs{F}^{\mathrm{e}}} \cdot {\bs{F}^{\mathrm{c}}}^{-T} \right) \colon \dot{\bs{F}} + {\bs{F}^{\mathrm{e}}}^T \cdot \frac{\partial \Psi^{\mathrm{e}}}{\partial \bs{F}^{\mathrm{e}}} \colon \dot{\bs{F}^{\mathrm{c}}} \cdot {\bs{F}^{\mathrm{c}}}^{-1} - \frac{\partial \Psi^{\mathrm{c}}}{\partial \chi} \dot{\chi} &\ge 0 \text{ .}
  \label{di2}
 \end{align}
\end{linenomath*}
Inequality \eqref{di2} is satisfied if the term in the brackets vanishes and thus the constitutive relation for the first Piola-Kirchhoff stress tensor is obtained
\begin{linenomath*}
\begin{equation}
  \bs{P} = \frac{\partial \Psi^{\mathrm{e}}}{\partial \bs{F}^{\mathrm{e}}} \cdot {\bs{F}^{\mathrm{c}}}^{-T} \text{ .}
  \label{Pi}
\end{equation}
\end{linenomath*}
By defining the remaining expressions in Eq. \eqref{di2} as the velocity gradient $\bs{L}^{\mathrm{c}}$ and the Mandel stress tensor $\bs{M}$
\begin{linenomath*}
\begin{equation}
   \bs{L}^{\mathrm{c}} := \dot{\bs{F}^{\mathrm{c}}} \cdot {\bs{F}^{\mathrm{c}}}^{-1} \quad \text{and} \quad \bs{M} := {\bs{F}^{\mathrm{e}}}^T \cdot \frac{\partial \Psi^{\mathrm{e}}}{\partial \bs{F}^{\mathrm{e}}}
\end{equation}
\end{linenomath*}
the reduced dissipation inequality yields
\begin{linenomath*}
\begin{equation}
  D = \bs{M} \colon \bs{L}^{\mathrm{c}} - \frac{\partial \Psi^{\mathrm{c}}}{\partial \chi} \dot{\chi} \ge 0 \text{ .}
  \label{di3}
\end{equation}
\end{linenomath*}

\section{Derivation of evolution equations}
\label{sec2}

\subsection{Thermodynamic framework}
\label{mpdp}
An important part of constitutive modeling is the postulation of evolution laws for internal variables. In contrast to traditional phenomenological models, where direct assumptions for evolution equations are made according to experimental results, energetically based approaches are mostly applied in up-to-date models. Here, two alternative ways can be pursued. The principle of maximum dissipation postulates that the actual thermodynamic fluxes maximize the dissipation subjected to the subsidiary constraints \cite{onsager1931,hackl2008}. The present contribution, however, deals with the minimum principle of the dissipation potential \cite{PhysRev.97.1463}. In order to express this principle a general case is considered. Here, the Helmholtz energy $\Psi(\bs{\Gamma},\bs{\gamma})$ depends on a set of external variables $\bs{\Gamma} = \{ \Gamma_1, \Gamma_2, \ldots, \Gamma_{n_\mathrm{ex}} \} $ and on a set of internal variables $\bs{\gamma} = \{ \gamma_1, \gamma_2, \ldots, \gamma_{n_\mathrm{in}}  \} $, such that the rate of the Helmholtz energy is then given by
\begin{linenomath*}
\begin{equation}
  \dot{\Psi} = \frac{\partial \Psi}{\partial \bs{\Gamma}} \colon \dot{\bs{\Gamma}} + \frac{\partial \Psi}{\partial \bs{\gamma}} : \dot{\bs{\gamma}} \text{ .}
  \label{dpsi}
\end{equation}
\end{linenomath*}
The first term in Eq. \eqref{dpsi} is typically used to define constitutive laws as was done in Eqs. \eqref{di2} and \eqref{Pi}. In contrast, the second term in Eq. \eqref{dpsi} is used to define the dissipation as follows
\begin{linenomath*}
\begin{equation}
  D = - \frac{\partial \Psi}{\partial \bs{\gamma}} : \dot{\bs{\gamma}} = \bs{q}_{\bs{\gamma}} \colon \dot{\bs{\gamma}} \ge 0 \text{ .}
  \label{di4}
\end{equation}
\end{linenomath*}
The dissipation includes separate contributions due to the thermodynamic fluxes $\dot{\bs{\gamma}}$ and their conjugate pairs also known as thermodynamic driving forces
\begin{linenomath*}
\begin{equation}
  \bs{q}_{\bs{\gamma}} := - \partial \Psi / \partial \bs{\gamma} \text{ .}
  \label{qgam1}
\end{equation}
\end{linenomath*}
Finally, the minimum principle of the dissipation potential is expressed by
\begin{linenomath*}
\begin{equation}
  \min \{ \mathcal{L} = \dot{\Psi} + \Delta \, | \, \dot{\bs{\gamma}} \} \text{ .}
  \label{min}
\end{equation}
\end{linenomath*}
This principle enables the derivation of the evolution laws for the internal variables $\dot{\bs{\gamma}}$ by minimizing the Lagrangian $\mathcal{L}$ composed of the Helmholtz energy rate $\dot{\Psi}$ and the dissipation potential $\Delta$. The stationary point of the Lagrangian 
\begin{linenomath*}
\begin{equation}
 \mathcal{L} =  \frac{\partial \Psi}{\partial \bs{\Gamma}} \colon \dot{\bs{\Gamma}} + \frac{\partial \Psi}{\partial \bs{\gamma}} : \dot{\bs{\gamma}} + \Delta(\dot{\bs{\gamma}})
 \label{min2}
\end{equation}
\end{linenomath*}
is sought, which implies that its first derivative with respect to the fluxes $\dot{\bs{\gamma}}$ has to be equal to zero
\begin{linenomath*}
\begin{equation}
  \frac{\partial \mathcal{L}}{\partial \dot{\bs{\gamma}}} = \frac{\partial \Psi}{\partial \bs{\gamma}} + \frac{\partial \Delta}{\partial \dot{\bs{\gamma}}} = 0 \quad \Rightarrow \quad \frac{\partial \Delta}{\partial \dot{\bs{\gamma}}} = - \frac{\partial \Psi}{\partial \bs{\gamma}} \text{ .}
  \label{min3}
\end{equation}
\end{linenomath*}
According to Eqs. \eqref{qgam1} and \eqref{min3} the driving forces are expressed as derivative of the dissipation potential
\begin{linenomath*}
\begin{equation}
  \bs{q}_{\bs{\gamma}} = \frac{\partial \Delta}{\partial \dot{\bs{\gamma}}} \text{ .}
  \label{qgam2}
\end{equation}
\end{linenomath*}
The set-up \eqref{min}-\eqref{qgam2} is a generic procedure which is now used to derive equations driving the microstructure evolution in the case of the SIC. For this purpose, two assumptions are introduced in Sects. \ref{assumption} and \ref{ass_dis}.

\subsection{Assumption for coupling $\bs{F}^{\mathrm{c}}$ and $\chi$}
\label{assumption}
In the following, a model is chosen where the evolution of the regularity is influenced by the direction of the stretch. Crystallites are observed to be well oriented with their fiber axes parallel to the stretch direction \cite{doi:10.1063/1.1746537}. For this purpose, a coupling of the evolution of $\bs{F}^{\mathrm{c}}$ with the evolution of the regularity $\chi$ is introduced as follows
\begin{linenomath*}
\begin{equation}
   \bs{L}^{\mathrm{c}} = \dot{\bs{F}^{\mathrm{c}}} \cdot {\bs{F}^{\mathrm{c}}}^{-1} = k \, \dot{\chi} \, \bs{N}^{\mathrm{dev}} \text{ .}
   \label{coup_n1}
\end{equation}
\end{linenomath*}
Here, symbol $k$ denotes a positive proportionality constant between the rates of internal variables $\bs{F}^{\mathrm{c}}$ and $\chi $. 
The proposed relationship corresponds to the unfilled polymers being nearly incompressible materials such that the evolution direction $\bs{N}^{\mathrm{dev}}$ only depends on the deviatoric part of Mandel stress $\bs{M}^{\mathrm{dev}}$
\begin{linenomath*}
\begin{equation}
  \bs{N}^{\mathrm{dev}} = \frac{\bs{M}^{\mathrm{dev}}}{\lVert \bs{M}^{\mathrm{dev}} \rVert} \text{ ,} \quad \bs{M}^{\mathrm{dev}} = \bs{M} - \frac{\mathrm{tr}(\bs{M})}{3} \bs{I} \text{ .}
  \label{coup_n2}
\end{equation}
\end{linenomath*}
In the case of filled polymers, the volume changes play a more significant role such that  the coupling condition has to incorporate the total Mandel stress tensor instead of its deviatoric part
\begin{linenomath*}
\begin{equation}
   \bs{L}^{\mathrm{c}} = \dot{\bs{F}^{\mathrm{c}}} \cdot {\bs{F}^{\mathrm{c}}}^{-1} = k \, \dot{\chi} \, \bs{N} \text{ ,} \quad \bs{N} = \frac{\bs{M}}{\lVert \bs{M} \rVert} \text{ .}
   \label{coup_n3}
\end{equation}
\end{linenomath*}
It should also be pointed out that an alternative  form of definition \eqref{coup_n1} is possible since the proposed material model is isotropic.  In this case, the elastic right Cauchy-Green tensor $\bs{C}^{\mathrm{e}}$ and the Mandel stresses $\bs{M}$ are coaxial as their spectral decompositions show 
\begin{linenomath*}
\begin{equation}
  \bs{C}^{\mathrm{e}} = \sum_{i=1}^3 C_i \, \bs{n}_i \otimes \bs{n}_i \text{ ,} \quad \bs{M} = \sum_{i=1}^3 M_i \, \bs{n}_i \otimes \bs{n}_i \text{.}
\end{equation} 
\end{linenomath*}
In the previous equations, $\bs{n}_i$ represent the eigenvectors and $C_i$, $M_i$ are the eigenvalues of the corresponding tensors.  Accordingly, $\bs{N}^{{\mathrm{dev}}}$ can also be expressed as %
\begin{linenomath*}
\begin{equation}
\bs{N}^{{\mathrm{dev}}} = \frac{\bs{C}^{\mathrm{e,dev}}}{\lVert \bs{C}^{\mathrm{e,dev}} \rVert} 
\text{ ,} \quad \bs{C}^{\mathrm{e,dev}} = \bs{C} - \frac{\mathrm{tr}(\bs{C})}{3} \bs{I} \text{ ,}
  \label{coup_n2a}
\end{equation}
\end{linenomath*}
 which directly correspond to the physical observation that crystal orientation depends on stretches. Both formulations  \eqref{coup_n2} and  \eqref{coup_n2a} are equivalent, however, the definition in terms of $\bs{M}^{\mathrm{dev}}$ is more appropriate  for later derivations.

By inserting coupling condition \eqref{coup_n1} into inequality \eqref{di3}, the reduced dissipation reads 
\begin{linenomath*}
\begin{equation}
  D = k \, \dot{\chi} \, \bs{M} \colon \frac{\bs{M}^{\mathrm{dev}}}{\lVert \bs{M}^{\mathrm{dev}} \rVert} - \frac{\partial \Psi^{\mathrm{c}}}{\partial \chi} \dot{\chi} = \left( k \, \lVert \bs{M}^{\mathrm{dev}} \rVert - c \right) \dot{\chi} \text{ .}
  \label{di_n}
\end{equation}
\end{linenomath*}
The only remaining internal variable is the regularity $\chi$ such that the dissipation inequality can be written in analogy to Eq. \eqref{di4}
\begin{linenomath*}
\begin{equation}
  D = q_{\chi} \, \dot{\chi} \ge 0 \text{ ,}
  \label{di_n2}
\end{equation}
\end{linenomath*}
where $q_{\chi}$ is the driving force
\begin{linenomath*}
\begin{equation}
  q_{\chi} := k \, \lVert \bs{M}^{\mathrm{dev}} \rVert - c \text{ .}
  \label{qchi}
\end{equation}
\end{linenomath*}
Accordingly
\begin{linenomath*}
\begin{equation}
  \dot{q}_{\chi} =  k \, \dot{\overline{\lVert \bs{M}^{\mathrm{dev}} \rVert}}
  \label{dqchi}
\end{equation}
\end{linenomath*}
determines the rate of the driving force corresponding to the chain regularity. Obviously, the sign of the rate of the driving force \eqref{dqchi} distinguishes the loading stage ($\dot{q}_{\chi} \ge 0$) and the unloading stage ($\dot{q}_{\chi} < 0$) at a single point.

\subsection{Assumption for the dissipation potential}
\label{ass_dis}
The procedure described in Sect. \ref{mpdp} requires the dissipation potential depending on the rate of the internal variable $\Delta(\dot{\chi})$ to be postulated. For this purpose, experimental results (Fig. \ref{fig1} b) are taken into account. Here, no development of crystalline regions is observed for the increasing load up to point A ($\lambda_A = 4.3$). The crystalline regions are built after $\lambda_A$ is exceeded and their volume fraction rises gradually. This motivates the following choice for the dissipation potential
\begin{linenomath*}
\begin{equation}
  \Delta = \left( A + B \right) \left\vert \dot{\chi} \right\vert \text{ ,}
  \label{di5}
\end{equation}
\end{linenomath*}
where $A$ denotes the crystallization limit and parameter $B$ determines the change of the crystallization limit depending on stresses applied. The evolution law for this parameter is coupled to the evolution of the regularity according to the expression 
\begin{linenomath*}
\begin{equation}
  \dot{B} = \frac{b}{f(\chi)} \left\vert \dot{\chi} \right\vert \text{ ,}
  \label{dB2}
\end{equation}
\end{linenomath*}
where $b$ is a material parameter controlling the velocity of the inelastic process during the loading and unloading phase
\begin{linenomath*}
\begin{equation}
  b = \begin{cases}
  b_1 & \text{if} \quad \dot{q}_{\chi} \geq 0 \text{ ,} \\
  b_2 & \text{if} \quad \dot{q}_{\chi} < 0 \text{ .}
  \end{cases} 
  \label{dB3}
\end{equation}
\end{linenomath*}
Moreover, the additional condition $b_2\!>\!b_1\!>\!0$ stipulates that the regularity decrease during the unloading phase is slower than its growth during the loading phase. Following the same line, function $f(\chi)$ is introduced to control the regularity evolution depending on its current value. In addition, this function enables the simulation of a faster crystallization at points with a higher regularity. A possible choice for the function is
\begin{linenomath*}
\begin{equation}
  f(\chi) = \alpha -(\chi - \beta)^2 \text{ ,} \quad f(\chi) > 0 \text{ ,}
  \label{dB4}
\end{equation}
\end{linenomath*}
where $\alpha$ and $\beta$ are material constants.

\subsection{Evolution equation for the regularity $\chi$}
\label{evo_eq}
\begin{figure}[h]
  \centering
  \psfrag{x}{$\dot{\chi}$}
  \psfrag{d}{$\Delta$}
  \psfrag{p}{$\Delta = (A+B)\left\vert\dot{\chi}\right\vert$}
  \includegraphics[width=0.55\textwidth]{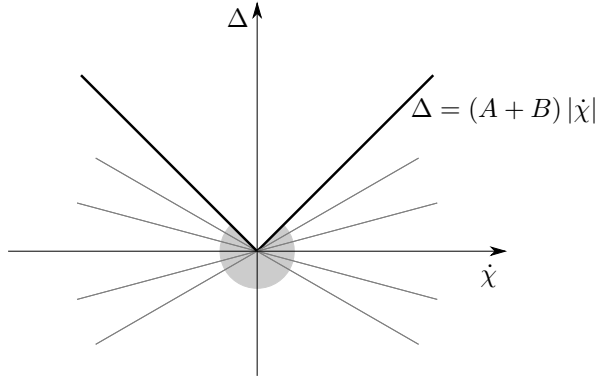}
  \caption{The subdifferential of the dissipation potential with the discontinuity at $\dot{\chi} = 0$.}
  \label{subdiff}
\end{figure}

In order to derive the evolution equation for the regularity $\chi$, the minimization problem \eqref{min} is considered. For this purpose, the rate of the Helmholtz energy is written based on explanations in Sects. \ref{thermo} and \ref{assumption} as
\begin{linenomath*}
\begin{equation}
  \dot{\Psi} = \bs{P} \colon \dot{\bs{F}} - q_{\chi} \, \dot{\chi}
  \label{dpsi1}
\end{equation}
\end{linenomath*}
such that the Lagrangian (Eq. \eqref{min}) turns into
\begin{linenomath*}
\begin{equation}
  \mathcal{L} = \bs{P} \colon \dot{\bs{F}} - q_{\chi} \, \dot{\chi} + (A + B) \left\vert \dot{\chi} \right\vert \text{ .}
  \label{L}
\end{equation}
\end{linenomath*}
Furthermore, Eq. \eqref{qgam2} is used to define the thermodynamic force. However, the absolute value function in Eq. \eqref{di5} is not differentiable at $\dot{\chi} = 0$. Hence, the subdifferential of the dissipation potential $\partial \Delta(\dot{\chi})$ is specified according to the situation shown in Fig. \ref{subdiff}. Here, it holds that
\begin{linenomath*}
\begin{equation}
  q_{\chi} = \frac{\partial \Delta}{\partial \dot{\chi}} = \left( A + B \right) \mathrm{sgn} \left( \dot{\chi} \right) = \left( A + B \right) \frac{\dot{\chi}}{\left\vert\dot{\chi} \right\vert} \quad \text{for} \quad \dot{\chi} \neq 0 \text{ ,}
  \label{first_sd}
\end{equation}
\end{linenomath*}
whereas any value $\left\vert q_{\chi} \right\vert \le A + B$ can be a solution for $\dot{\chi} = 0$.
 \\
In summary, the obtained subdifferential has the form:
\begin{linenomath*}
\begin{subequations}
\begin{empheq}[left={\partial \Delta (\dot{\chi}) = \empheqlbrace}]{align}
    \left\{ q_{\chi}; \, \left\vert q_{\chi} \right\vert \le A + B \right\} & \text{for} \quad \dot{\chi} = 0 \text{ ,}
    \label{sd_a}\\
    q_{\chi} = \left( A + B \right) \frac{\dot{\chi}}{\left\vert\dot{\chi} \right\vert}  & \text{for} \quad \dot{\chi} \neq 0 \text{ .}
    \label{sd_b}
\end{empheq}
\end{subequations}
\end{linenomath*}
The stretched material during the loading phase undergoes elastic deformations as long as criterion \eqref{sd_a} is fulfilled. After exceeding the crystallization limit, the regularity $\chi$ starts to evolve, which is described by Eq. \eqref{sd_b}. The same condition holds for the complete unloading stage, where the degradation of crystalline regions occurs. This is achieved by introducing a shift of the driving force explained in Sect. \ref{load_unload}. \\ \\

Equation \eqref{sd_b} is a crucial part of the model. This condition yields the evolution of the regularity
\begin{linenomath*}
\begin{equation}
  \dot{\chi} = \frac{\left\vert\dot{\chi} \right\vert}{A + B} q_{\chi} \text{ ,}
  \label{ev1}
\end{equation}
\end{linenomath*}
which can also be written in the form
\begin{linenomath*}
\begin{equation}
  \dot{\chi} = \lambda \, q_{\chi} \text{ ,} \quad \lambda \ge 0 \text{ ,}
  \label{ev2}
\end{equation}
\end{linenomath*}
where $\lambda$ is the crystallization parameter. The determination of this parameter relies on the results of relation \eqref{first_sd}
\begin{linenomath*}
\begin{equation}
  \left\vert q_{\chi} \right\vert = A + B \quad \text{,} \quad q_{\chi}^2 = \left( A + B \right)^2 \text{ ,} \quad q_{\chi} \, \dot{q}_{\chi} = \left( A + B \right) \dot{B} \text{ .}
  \label{yc}
\end{equation}
\end{linenomath*}
The insertion of Eqs. \eqref{qchi}, \eqref{dqchi}, \eqref{dB2} and \eqref{ev2} into Eq. \eqref{yc}c reads
\begin{linenomath*}
\begin{equation}
  \left( k \lVert \bs{M}^{\mathrm{dev}} \rVert - c \right) k \,  \dot{\overline{\lVert \bs{M}^{\mathrm{dev}} \rVert}} = \left( A + B \right)^2\frac{b}{f(\chi)} \lambda \text{ ,}
\end{equation}
\end{linenomath*}
such that the crystallization parameter turns into
\begin{linenomath*}
\begin{equation}
  \lambda = \frac{f(\chi) \left( k \lVert \bs{M}^{\mathrm{dev}} \rVert - c \right) k \, \dot{\overline{\lVert \bs{M}^{\mathrm{dev}} \rVert}}}{b \left( A + B \right)^2} \ge 0 \text{ .}
  \label{cp}
\end{equation}
\end{linenomath*}
The non-negativity of the crystallization parameter now requires special consideration. All constants in the denominator are positive, such that the sign of $\lambda$ is determined by the sign of the numerator, where function $f(\chi)$ is also positive. Consequently, the non-negativity of $\lambda$ implies that $q_{\chi} = k \lVert \bs{M}^{\mathrm{dev}} \rVert - c$ and $\dot{q}_{\chi} = k \, \dot{\overline{\lVert \bs{M}^{\mathrm{dev}} \rVert}}$ have the same sign, which is achieved by the suitable choice of constant $c$. Finally, the insertion of Eq. \eqref{ev2} into dissipation \eqref{di_n2} proves that the dissipation inequality
\begin{linenomath*}
\begin{equation}
  D = \lambda \, q_{\chi}^2 \ge 0
  \label{di_n3}
\end{equation}
\end{linenomath*}
is fulfilled due to the non-negativity of $\lambda$.

\section{Numerical implementation of the SIC model}
\label{sec3}
In order to numerically solve a boundary value problem depending on effects of SIC, the standard steps typical of the FEM implementation in the case of nonlinear materials and large deformations are performed. These steps deal with the  derivation of the residual and of the stiffness matrix as explained in Appendix A.

\subsection{Time discretization of evolution equations}
\label{td_eq}
The numerical implementation of evolution equations for hardening parameter $B$ and internal variable $\chi$ which are derived in Sect. \ref{sec2} requires discretization in time. To this end, the present contribution approximates derivatives by the forward differences, which leads to an explicit integration scheme. The evolution law \eqref{dB2} for parameter B is then expressed by
\begin{linenomath*}
\begin{equation}
  B_{n+1} = B_n + \frac{b}{f(\chi_n)} \left\vert \chi_{n+1} - \chi_n \right\vert \text{ ,}
  \label{dis_B}
\end{equation}
\end{linenomath*}
where subscript $n+1$ denotes values at current time step and subscript $n$ denotes values at previous time step. In addition, the following notation has been used for the numerical approximation of the time derivative
\begin{linenomath*}
\begin{equation}
  \dot{M}^{\mathrm{dev}}_n := \frac{\lVert \bs{M}^{\mathrm{dev}}_n \rVert - \lVert \bs{M}^{\mathrm{dev}}_{n-1} \rVert}{\Delta t} \text{ ,}
  \label{sgnm}
\end{equation} 
\end{linenomath*}
such that the explicit integration of the regularity evolution \eqref{ev2} is given by
\begin{linenomath*}
\begin{align}
  & \chi_{n+1} = \chi_n + \Delta \lambda \, q_{\chi \, n} \text{ ,} \\
  & \Delta \lambda = \frac{\Delta t \, f(\chi_n) \left( k \, \lVert \bs{M}^{\mathrm{dev}}_n \rVert -c \right) k \, \dot{M}^{\mathrm{dev}}_n}{b \left( A + B_n \right)^2} \text{ ,} \\
  & q_{\chi\, n} = k \lVert \bs{M}^{\mathrm{dev}}_n \rVert - c \text{ .}
  \label{dis_chi}
\end{align}
\end{linenomath*}
The evaluation of the Mandel stresses in the previous expressions requires the time integration of tensor valued quantities, an issue which has to be considered more closely. For this purpose, the evolution law \eqref{coup_n1} is first rewritten in the form
\begin{linenomath*}
\begin{equation}
  \dot{\bs{F}^{\mathrm{c}}} = k \, \dot{\chi} \, \bs{N}^{\mathrm{dev}} \cdot \bs{F}^{\mathrm{c}} \text{ .}
  \label{evo_dis3}
\end{equation}
\end{linenomath*}
Furthermore, the differential equation \eqref{evo_dis3} is numerically solved by applying the exponential map \cite{de2011computational}
\begin{linenomath*}
\begin{equation}
  \bs{F}^{\mathrm{c}}_{n+1} = \mathrm{exp} \left( k \left( \chi_{n+1} - \chi_n \right) \bs{N}^{\mathrm{dev}}_n \right) \cdot \bs{F}^{\mathrm{c}}_n \text{ .}
   \label{dis_Fc}
\end{equation}
\end{linenomath*}
The contribution by Moler and Van Loan (2003) \cite{moler2003} discusses and compares various ways to compute the exponential of a second order tensor. However, although some of the methods are preferable to others, none is entirely satisfactory. The method used in the current approach goes back to the definition of the tensor exponential: the numerical solution is carried out by calculating a finite truncation of the Taylor series.

\subsection{Numerical simulation of the unloading phase}
\label{load_unload}
Experimental results (Fig. \ref{fig1} b) show that the loading stage is related to the regularity increase, whereas the degradation of crystalline regions occurs during the unloading stage. In the present model, this change is determined by evolution equation \eqref{ev2} and  by definition of the driving force \eqref{qchi}. The development/degradation of crystalline regions is controlled by the sign of the driving force due to the non-negativity of $\lambda$. A negative driving force during the unloading stage is achieved by  introducing  shift $c$ in Eq. \eqref{qchi}. This load dependent parameter is calculated  from the condition for the initial value of driving force $q_{\chi}^{\mathrm{un, in}}$ to coincide with the negative crystallization limit if the increment $B$ is set to zero:
\begin{linenomath*}
\begin{equation}
  q_{\chi}^{\mathrm{un, in}} = k \, \lVert \bs{M}^{\mathrm{dev}}_{\mathrm{end},\mathrm{ld}} \rVert - c = - A \quad \Rightarrow \quad c = A + k \lVert \bs{M}^{\mathrm{dev}}_{\mathrm{end},\mathrm{ld}} \rVert \text{ .}
  \label{q_sh}
\end{equation}
\end{linenomath*}
Here, $\bs{M}^{\mathrm{dev}}_{\mathrm{end},\mathrm{ld}}$ is the deviatoric Mandel stress tensor at the end of loading stage, superscript ``ld''  denotes  the loading stage, ``un'' the unloading stage and ``in'' an initial value.  The relationship \eqref{q_sh} can easily be generalized for a multicyclic test, where shifts for loading (ld) and unloading (un) stages of a cycle $i$ are defined as follows:
\begin{linenomath*}
\begin{align}
&c^{\mathrm{ld},1} = 0 \text{ ,} \quad  c^{\mathrm{ld},i} = - k \lVert \bs{M}^{\mathrm{dev}}_{\mathrm{end},\mathrm{un},i-1} \rVert \text{ ,} \quad i = 2, ... , n &&\text{(loading stage)} \text{ ,} \\
&c^{\mathrm{un},i} = k \lVert \bs{M}^{\mathrm{dev}}_{\mathrm{end},\mathrm{ld},i} \rVert + A \text{ ,} \quad i = 1, ... , n &&\text{(unloading stage)} \text{ .}
\end{align}
\end{linenomath*}
As previously mentioned, the increment $B$ is reset at each change  between the loading and unloading modes.

\section{Numerical examples}
\label{sec4}
\begin{figure}[h!]
 \psfrag{u}{\footnotesize $\bar{u}$}
 \psfrag{1}[b]{\footnotesize $l=100$ nm}
 \psfrag{2}[]{\footnotesize $l=100$ nm}
 \psfrag{e1}{\footnotesize $\bs{e}_1$}
 \psfrag{e2}[cc]{\footnotesize $\bs{e}_2$}
 \psfrag{t}[cc]{\footnotesize $t$ [s]}
 \psfrag{lam}[]{\footnotesize $\bar{\lambda}$}
 \centering
   (a)\subfloat{\includegraphics[width=0.40\textwidth]{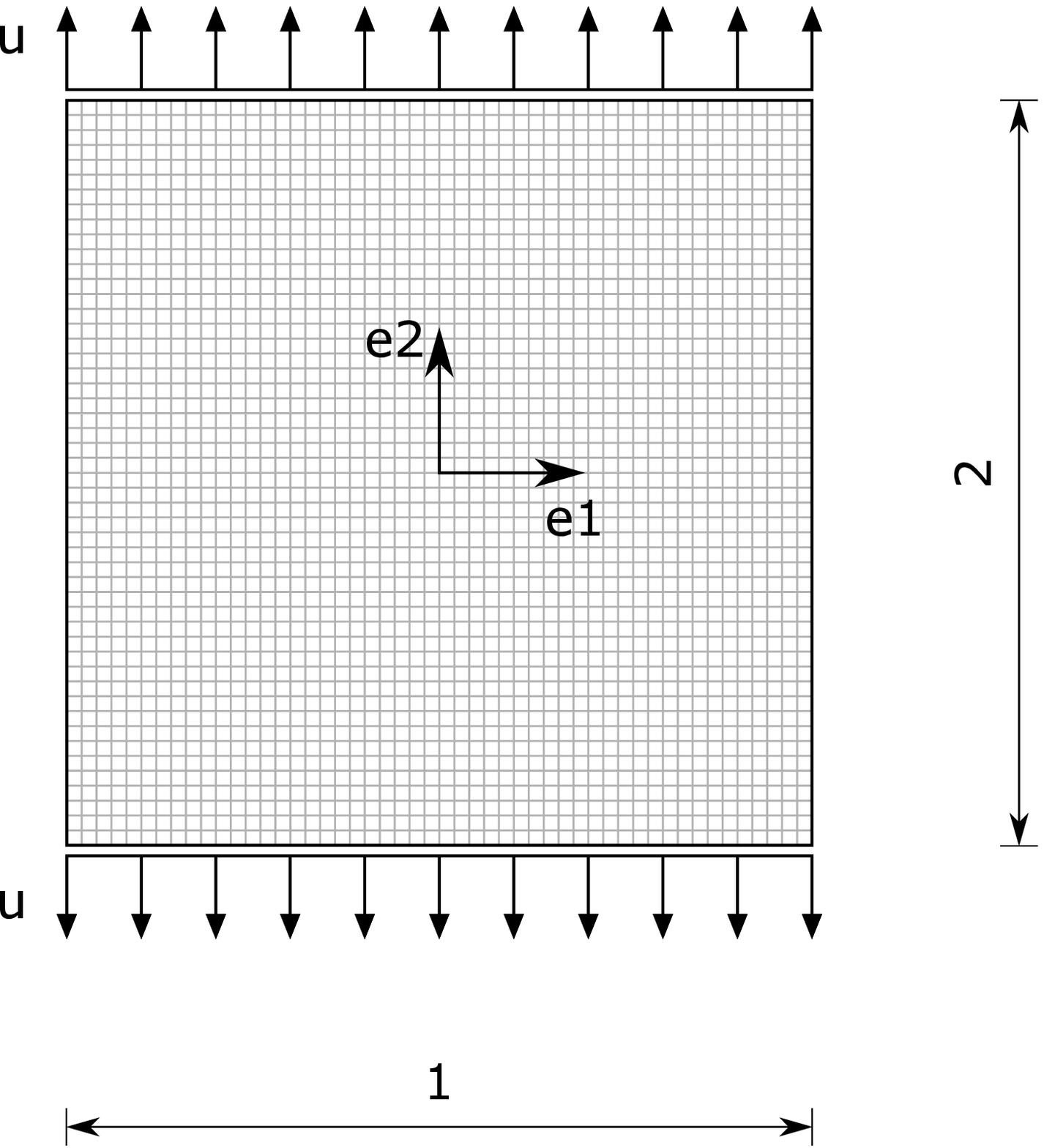}} \hspace{6mm}
  (b)\subfloat{\raisebox{8mm}{\includegraphics[width=0.45\textwidth]{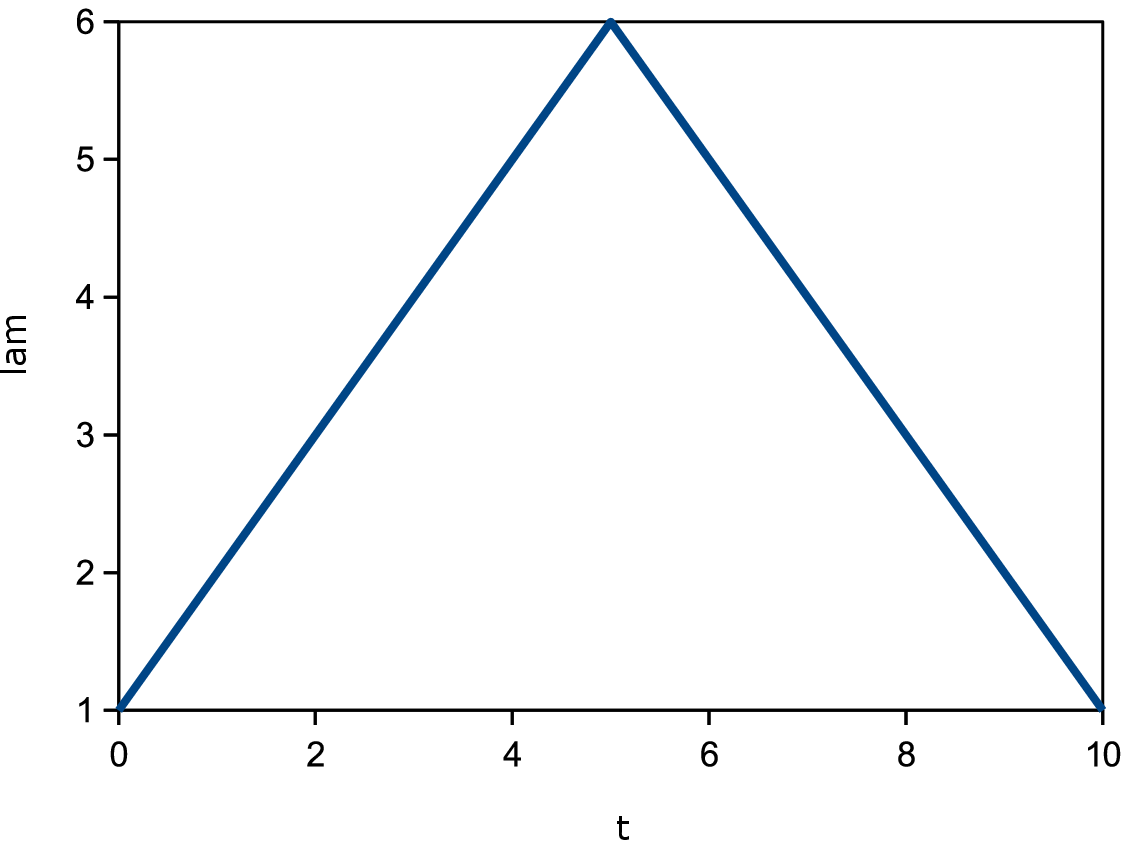}} }
 \caption{(a) Geometry and discretization of the sample with the prescribed vertical displacement $\bar{u}$ and side length $l$. (b) The applied stretch $\bar{\lambda} = (l+2\,\bar{u}) / l$ as a function of time.}
 \label{init}
\end{figure}

\begin{figure}[h!]
 \raggedright
 \hspace{0.4cm} $\bar{\lambda} = 5.15$ \hspace{2.55cm} $\bar{\lambda} = 5.40$ \hspace{2.55cm} $\bar{\lambda} = 5.71$ \\\vspace{2mm}
 \centering
 (a)\,\includegraphics[width=0.285\textwidth]{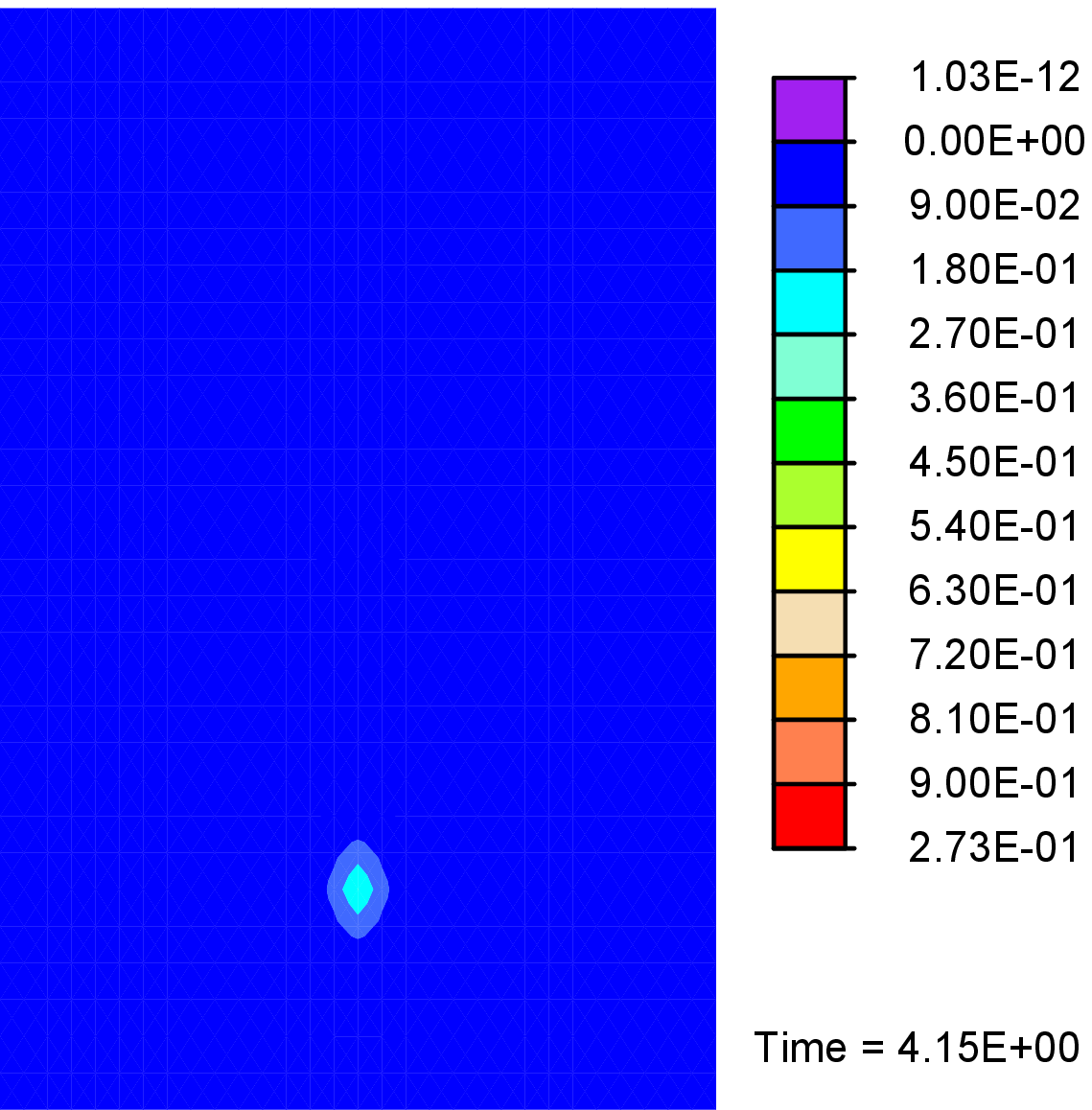}
 (b)\,\includegraphics[width=0.285\textwidth]{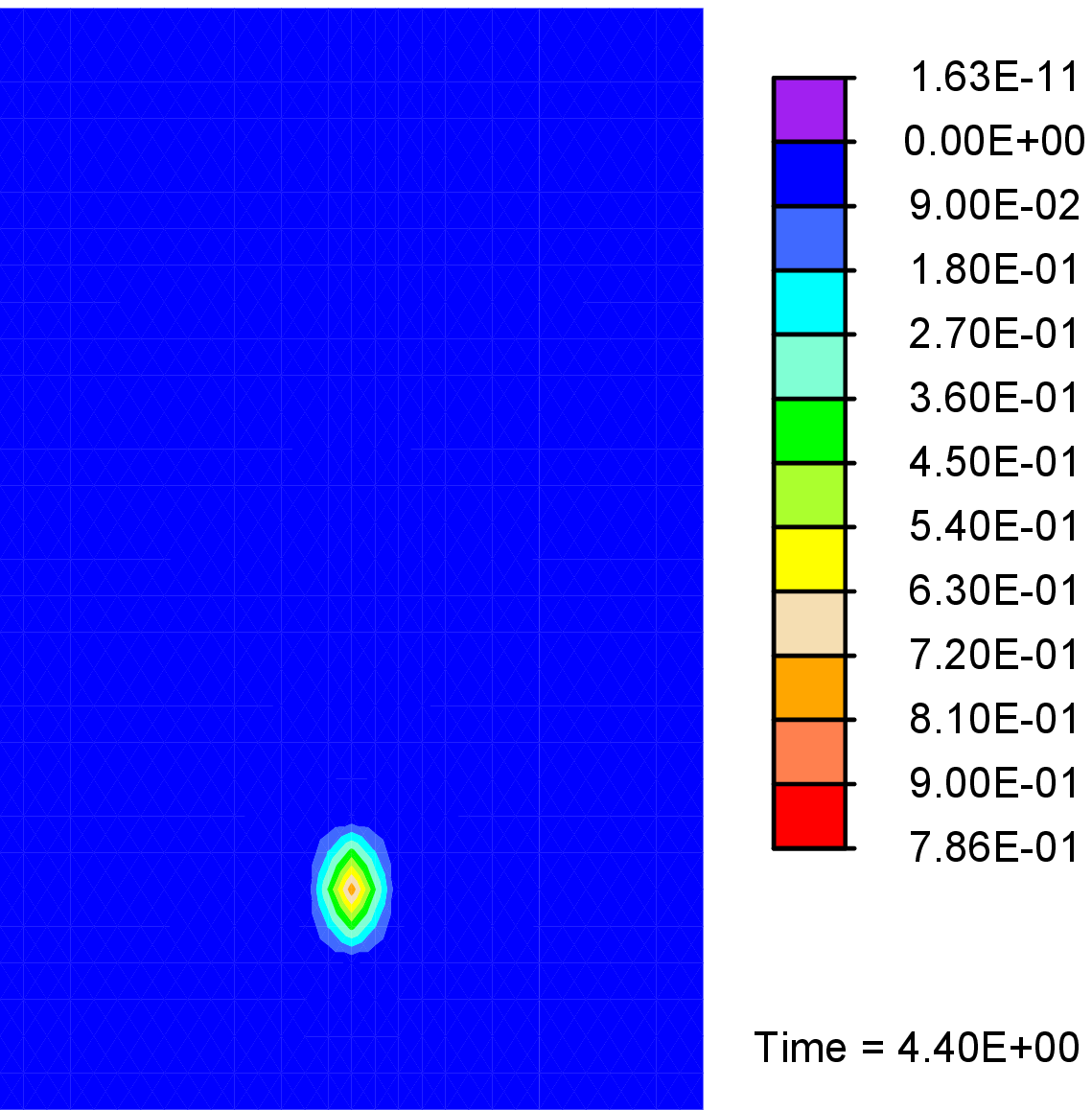}
 (c)\,\includegraphics[width=0.285\textwidth]{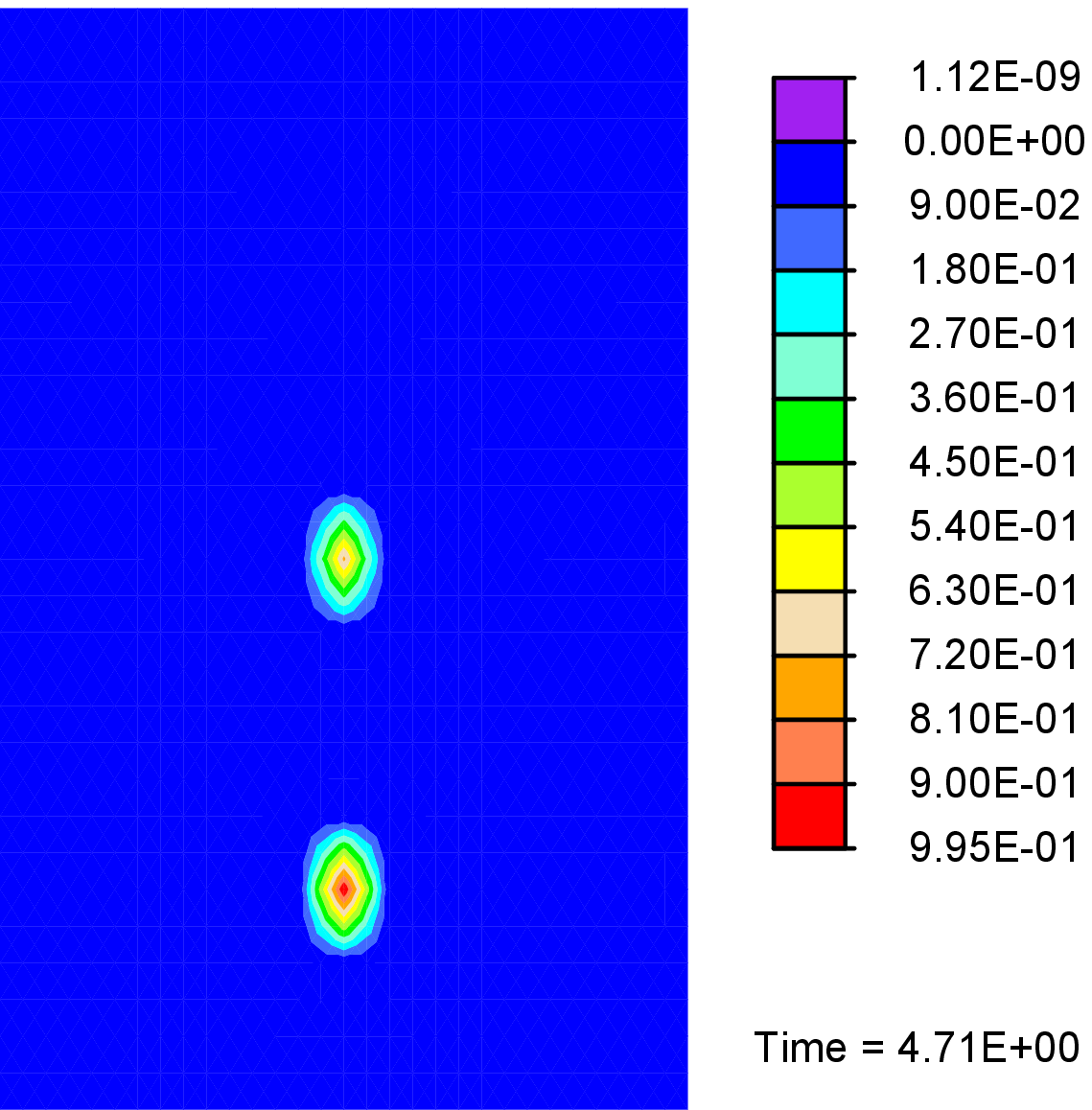} \\\vspace{5mm}
 \raggedright
 \hspace{0.4cm} $\bar{\lambda} = 6.00$ \hspace{2.55cm} $\bar{\lambda} = 5.77$ \hspace{2.55cm} $\bar{\lambda} = 5.26$ \\ \vspace{2mm}
 \centering
 (d)\,\includegraphics[width=0.285\textwidth]{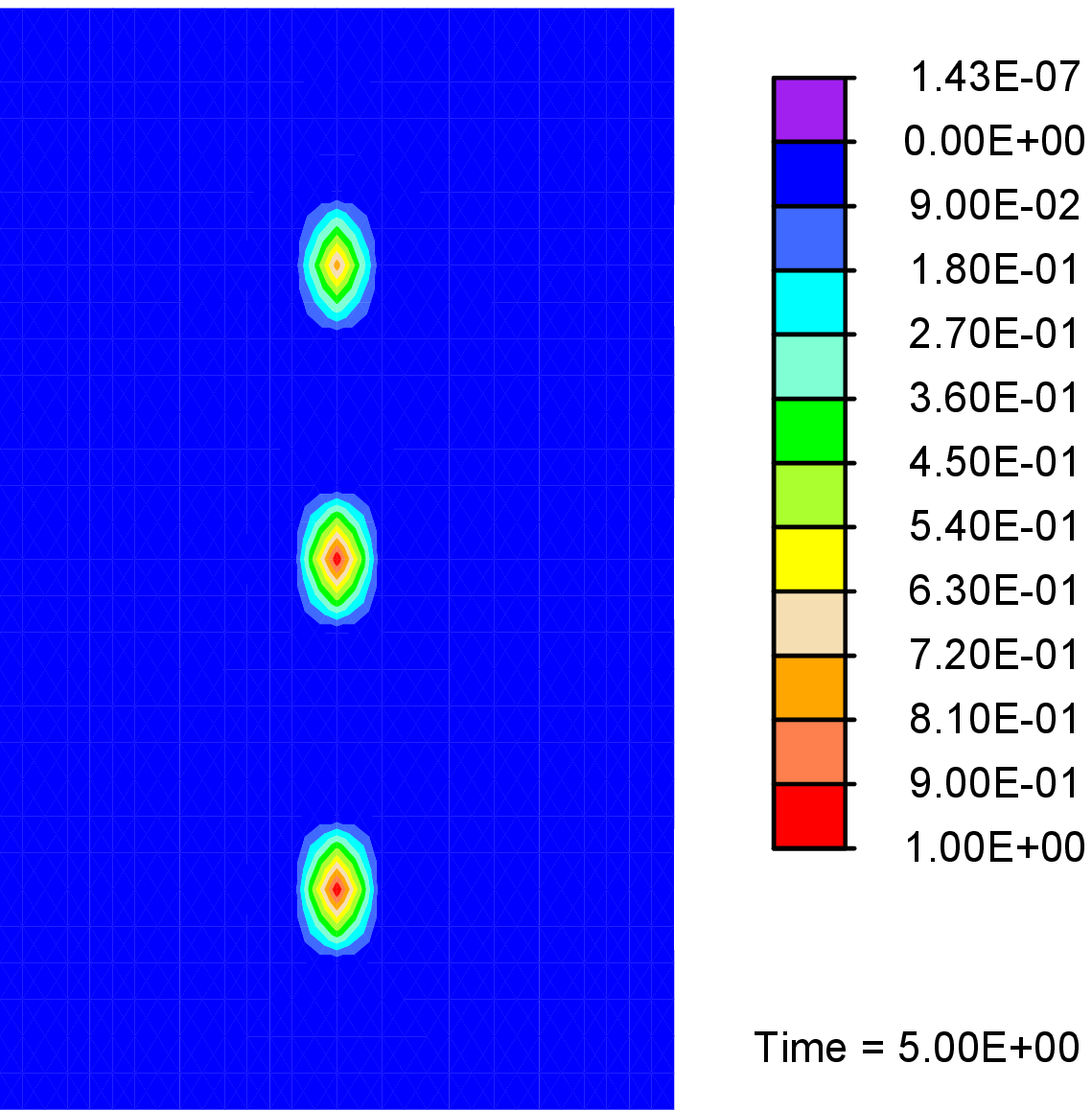}
 (e)\,\includegraphics[width=0.285\textwidth]{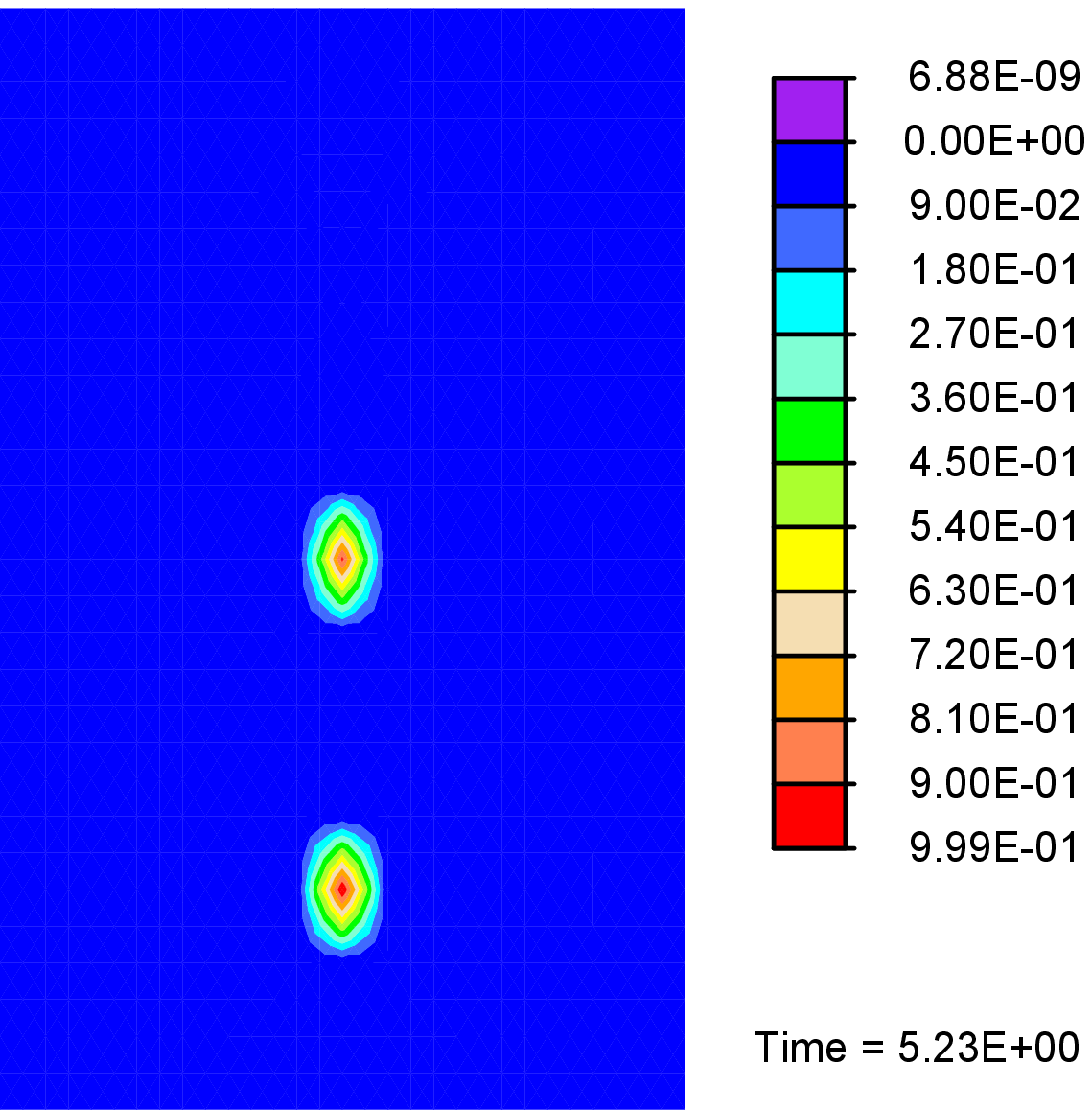}
 (f)\,\includegraphics[width=0.285\textwidth]{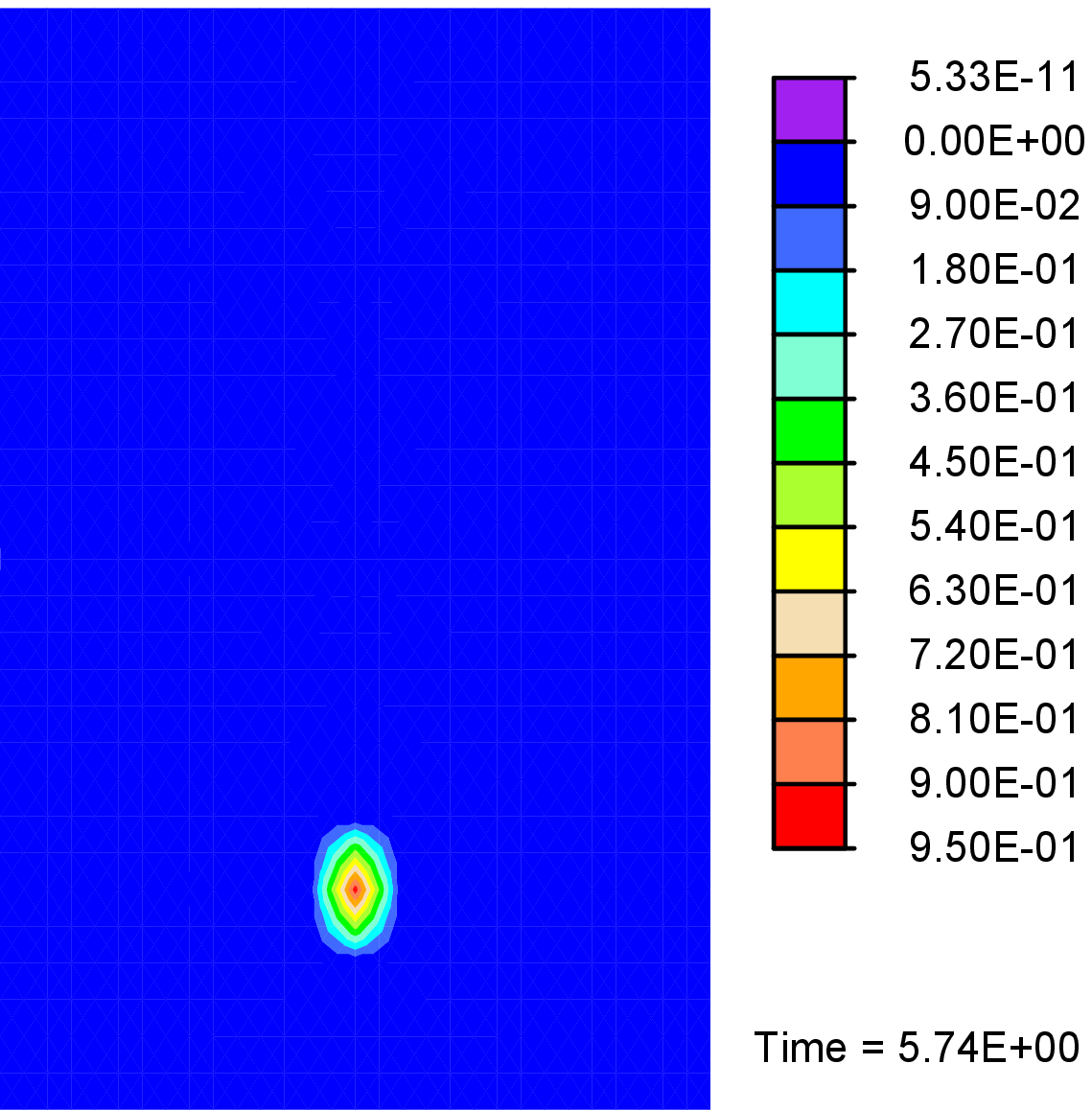}
 \caption{Simulation of the cyclic tension test on a sample with three nuclei. The initial regularities are $\chi_0 =$ \mbox{1E-8}, \mbox{1E-6} and \mbox{1E-4} (from top to bottom). (a)-(c) Snapshots of the microstructure during the loading phase. (d) State of the microstructure at the end of loading. (e)-(f) Snapshots of the microstructure during the unloading phase. The microstructure is shown in deformed configuration.}
 \label{test1}
\end{figure}

\begin{figure}[h!]
 \raggedright
 \hspace{0.4cm} $\bar{\lambda} = 5.14$ \hspace{2.55cm} $\bar{\lambda} = 6.00$ \hspace{2.55cm} $\bar{\lambda} = 4.54$ \\ \vspace{2mm}
 \centering
 (a)\,\includegraphics[width=0.285\textwidth]{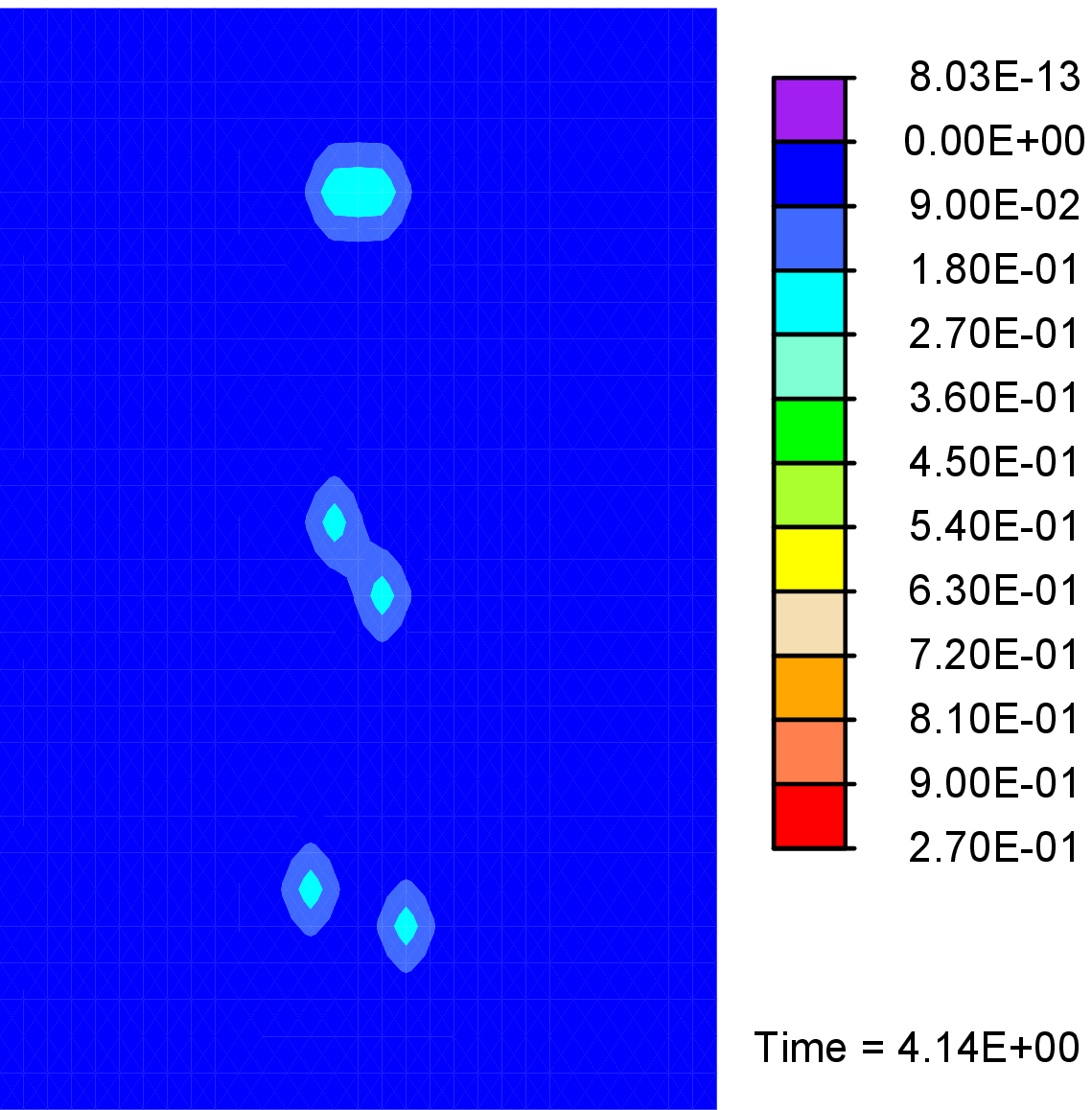}
 (b)\,\includegraphics[width=0.285\textwidth]{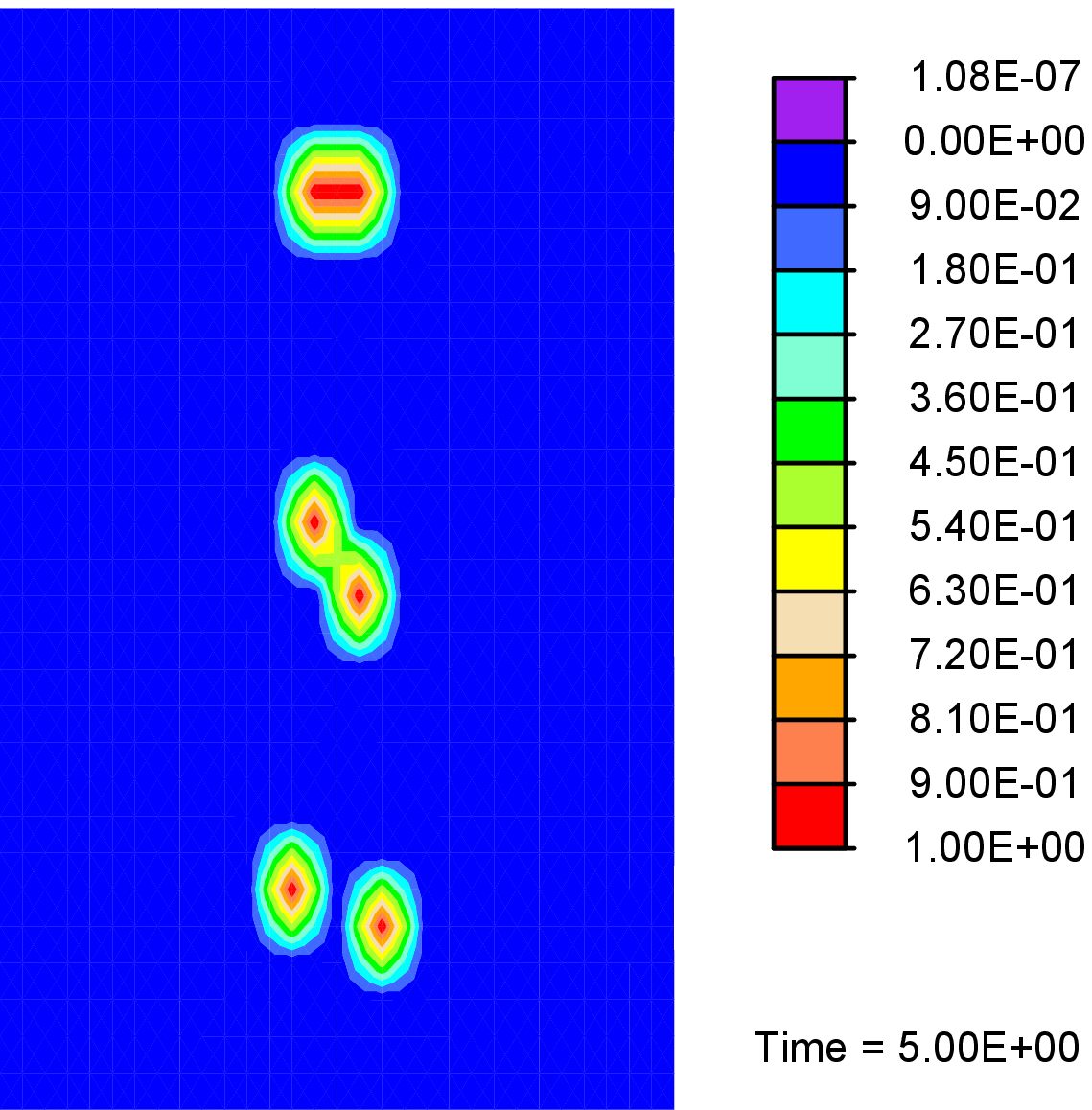}
 (c)\,\includegraphics[width=0.285\textwidth]{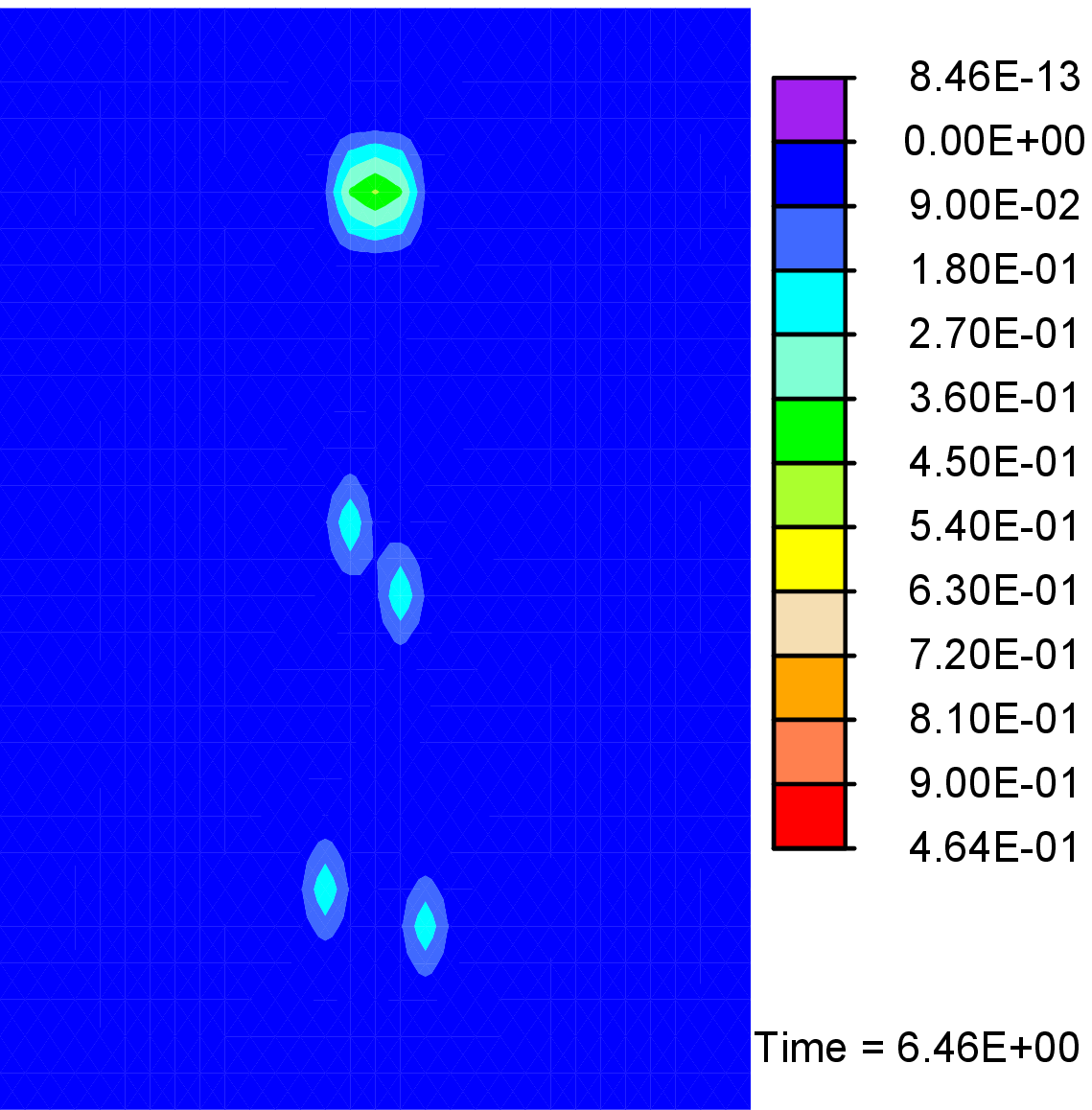} 
 \caption{Results of a cyclic tension test showing the influence of interaction and size of nuclei. (a) Snapshot of the microstructure during the loading phase. (b) State of the microstructure at the end of loading. (c) Snapshot of the microstructure during the unloading phase. The microstructure is shown in deformed configuration.}
 \label{test2}
\end{figure}

\begin{figure}[h!]
 \vspace{1mm}
 \raggedright
 \hspace{0.6cm} $\bar{\lambda} = 1.00$ \hspace{4.95cm} $\bar{\lambda} = 4.00$ \\ \vspace{2mm}
 \centering
 (a)\,\includegraphics[width=0.39\textwidth]{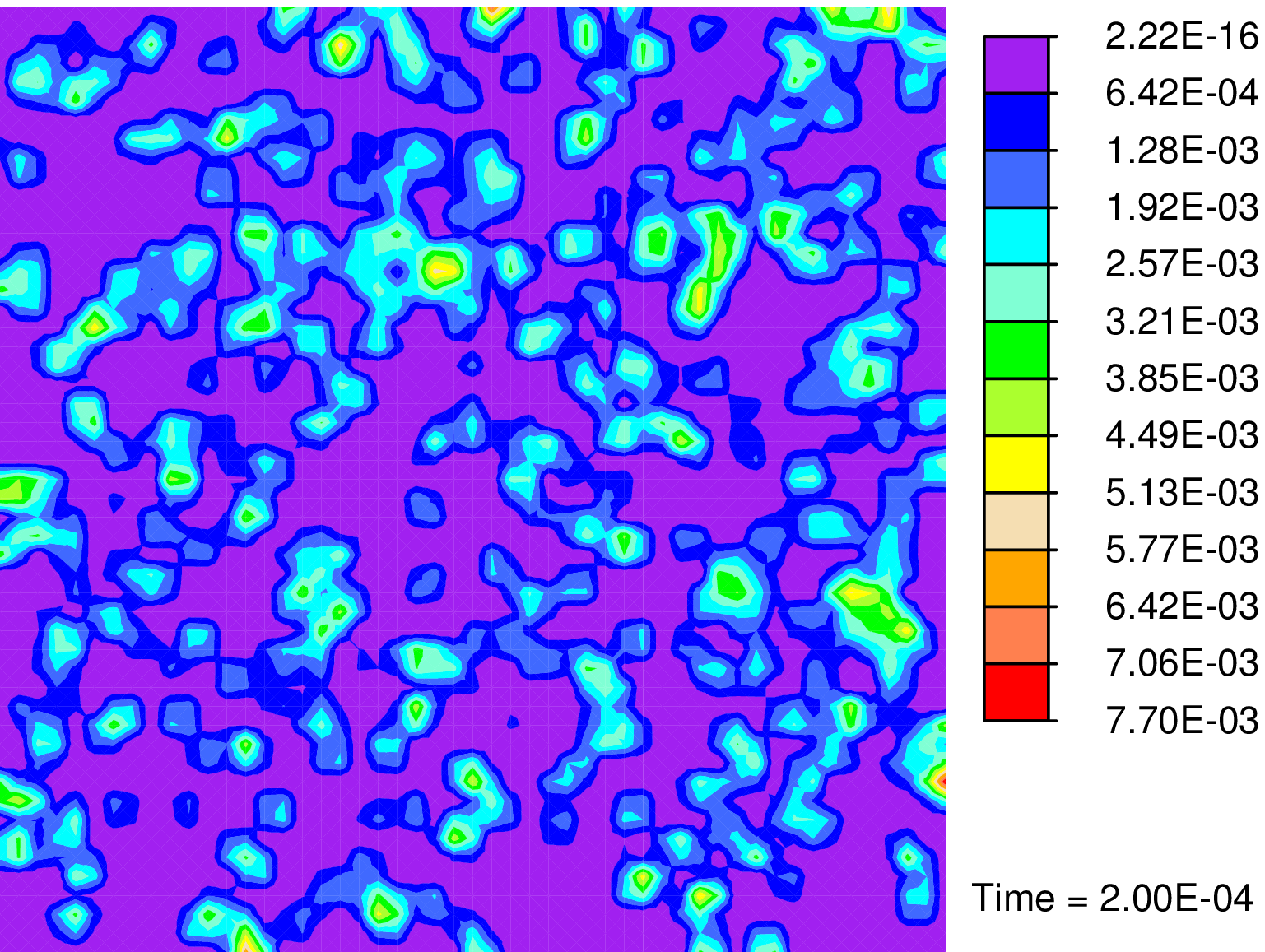} \hspace{1cm}
 (b)\,\includegraphics[width=0.39\textwidth]{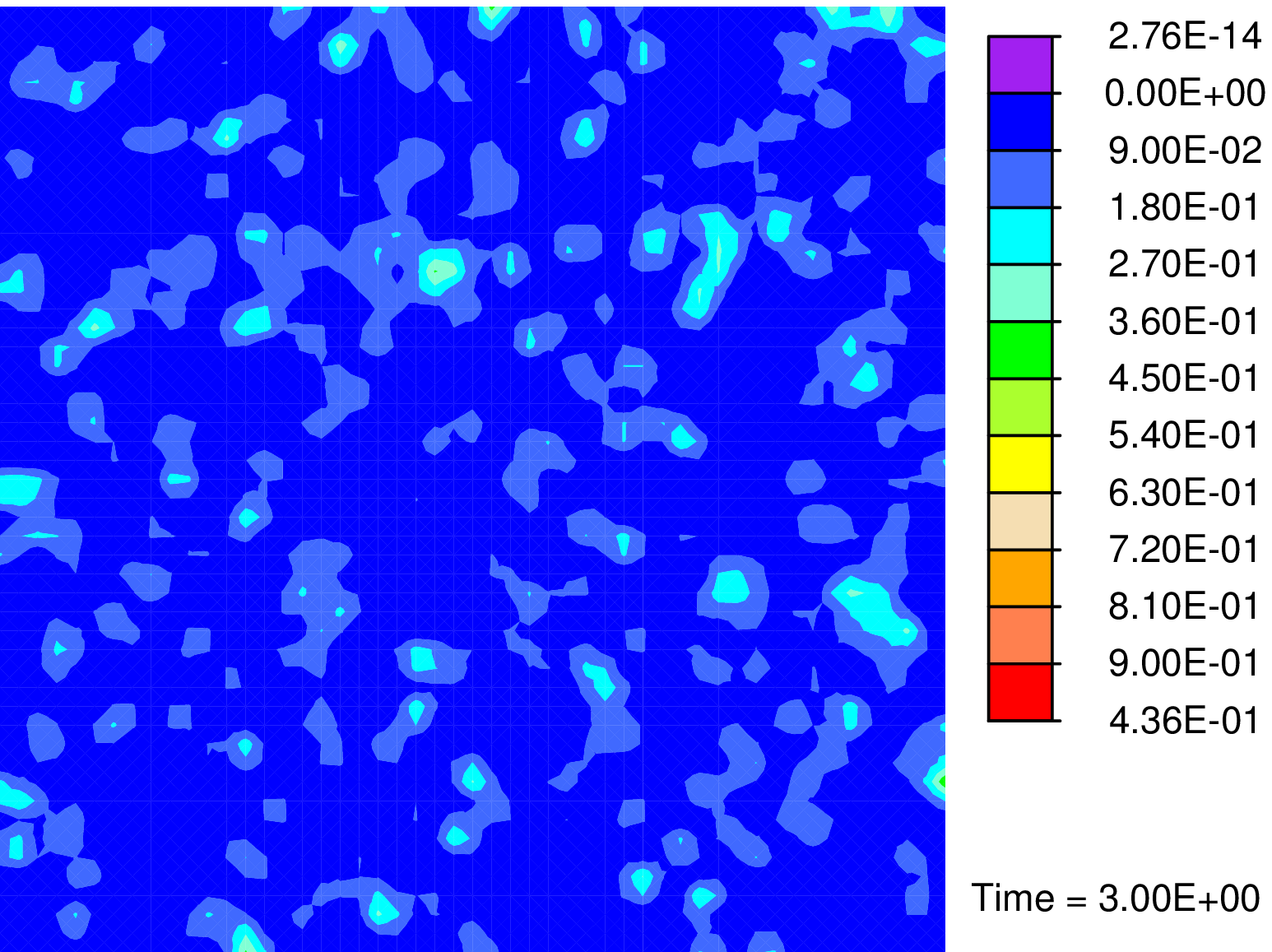} \\ \vspace{5mm}
  \raggedright
 \hspace{0.6cm} $\bar{\lambda} = 6.00$ \hspace{5cm} $\bar{\lambda} = 4.00$ \\ \vspace{2mm}
 \centering
 (c)\,\includegraphics[width=0.39\textwidth]{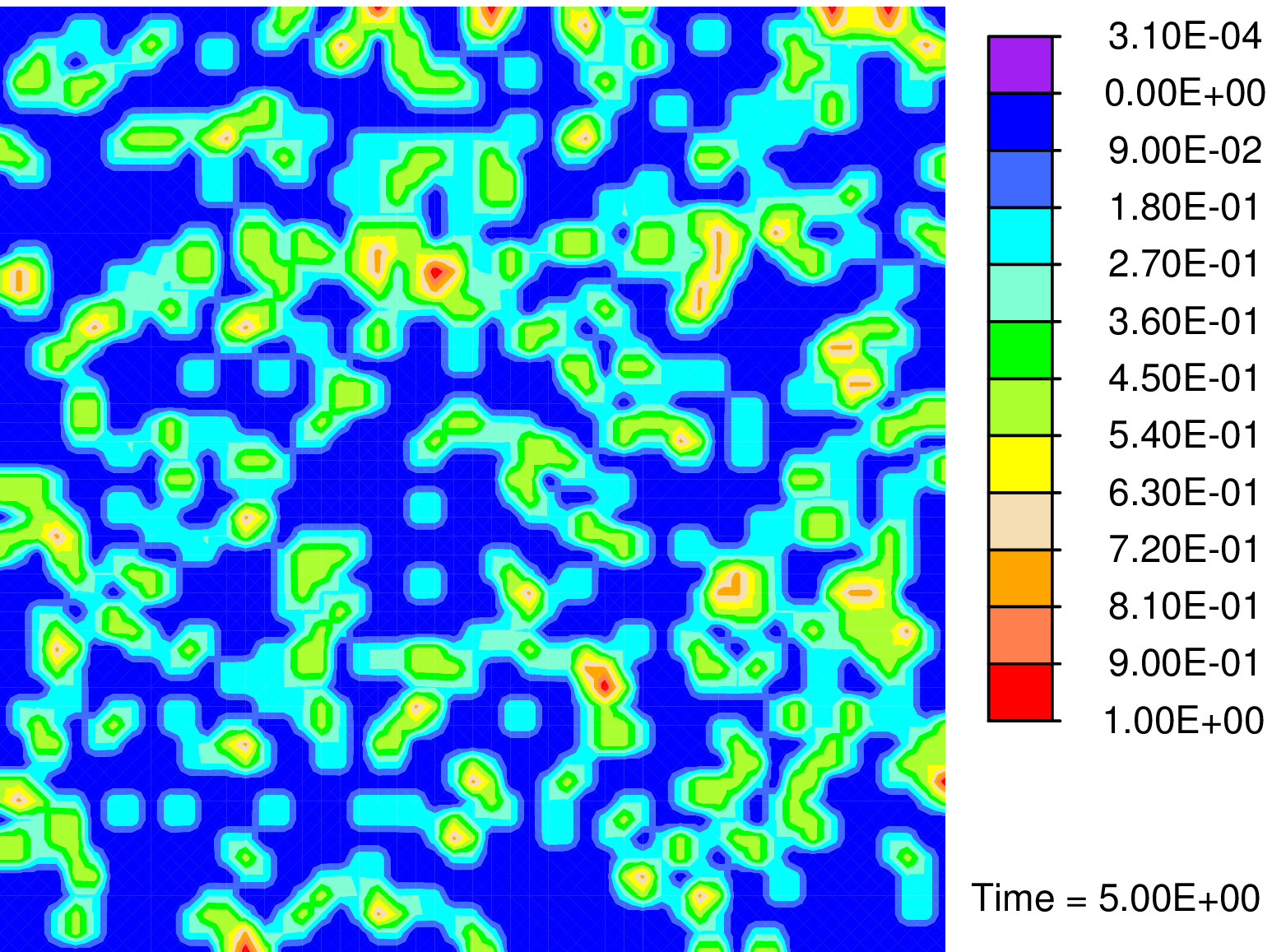} \hspace{1cm}
 (d)\,\includegraphics[width=0.39\textwidth]{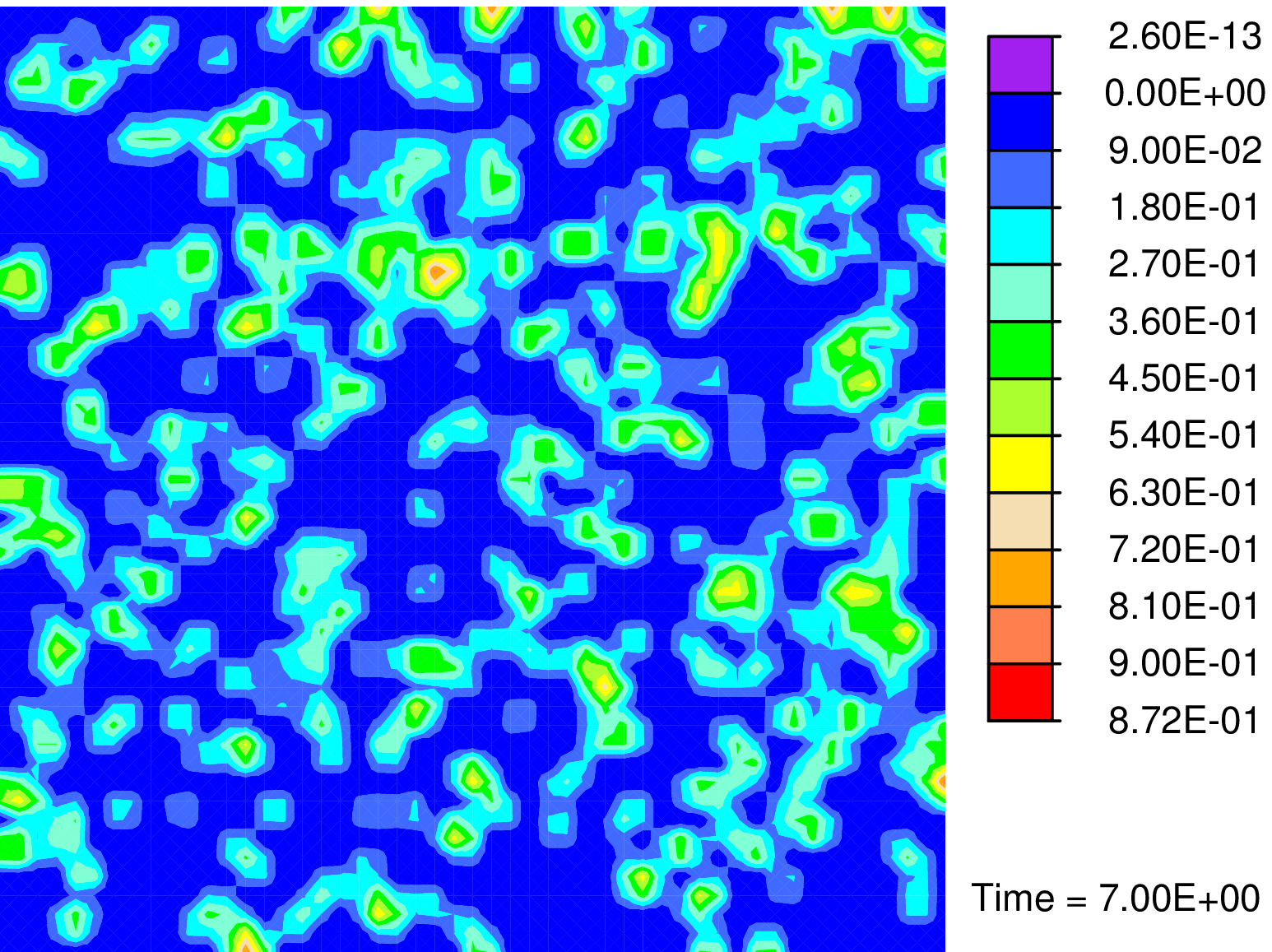}
 \begin{picture}(0,0)
   \put(-274,213){\line(1,1){35}}
   \put(-240,250){P}
 \end{picture}
 \caption{Cyclic tension test for a sample with a random initial microstructure. (a) Initial microstructure (type I). (b) Snapshot of the microstructure during the loading phase ($\bar{\lambda} = 4$). (c) State of the microstructure at the end of loading ($\bar{\lambda} = 6$). (d) Snapshot of the microstructure during the unloading phase ($\bar{\lambda} = 4$).}
 \label{test3}
\end{figure}

\begin{figure}[h!]
 \vspace{1mm}
 \raggedright
 \hspace{0.6cm} $\bar{\lambda} = 1.00$ \hspace{4.95cm} $\bar{\lambda} = 4.00$ \\ \vspace{2mm}
 \centering
 (a)\,\includegraphics[width=0.39\textwidth]{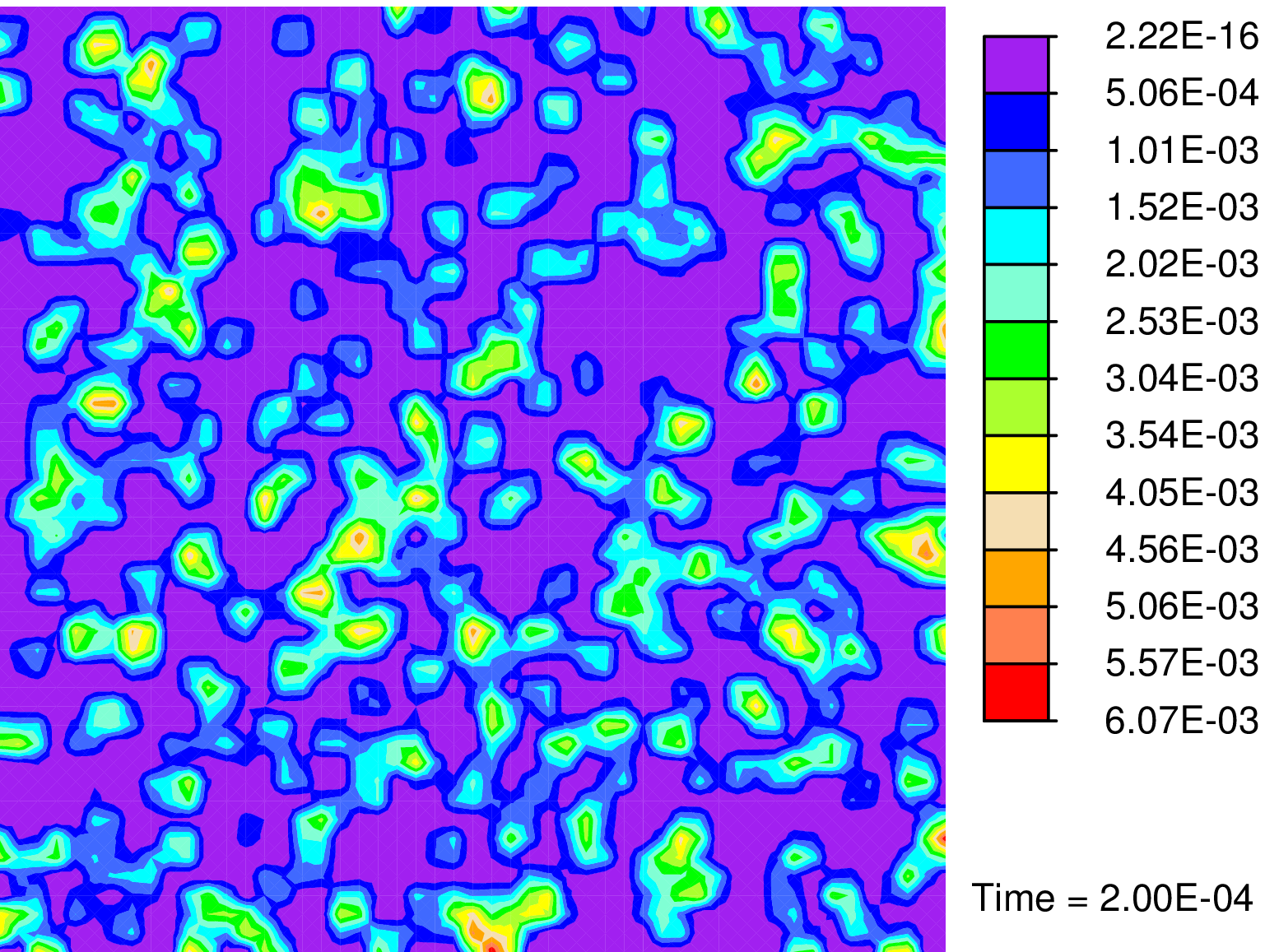} \hspace{1cm}
 (b)\,\includegraphics[width=0.39\textwidth]{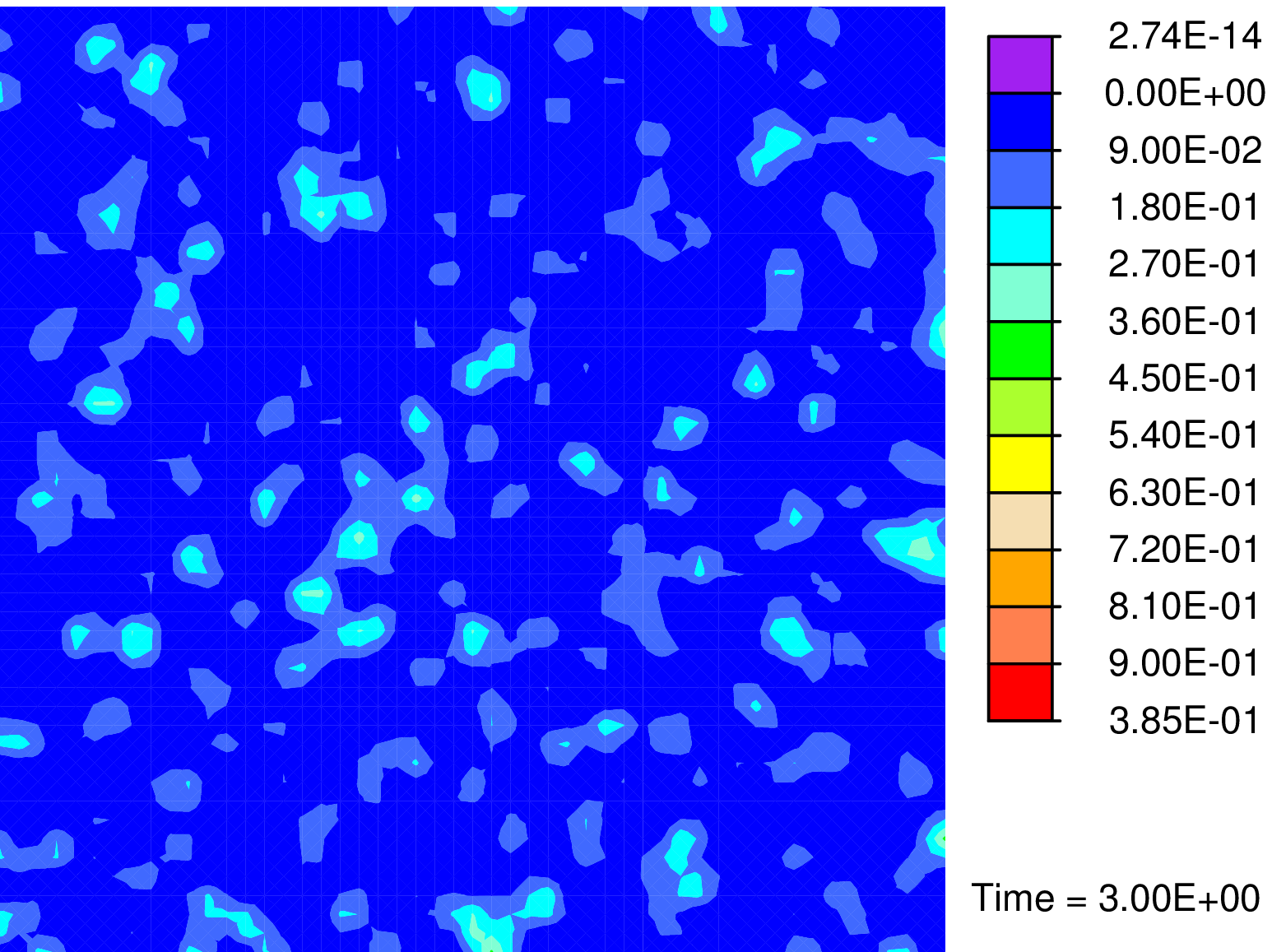} \\ \vspace{5mm}
  \raggedright
 \hspace{0.6cm} $\bar{\lambda} = 6.00$ \hspace{5cm} $\bar{\lambda} = 4.00$ \\ \vspace{2mm}
 \centering
 (c)\,\includegraphics[width=0.39\textwidth]{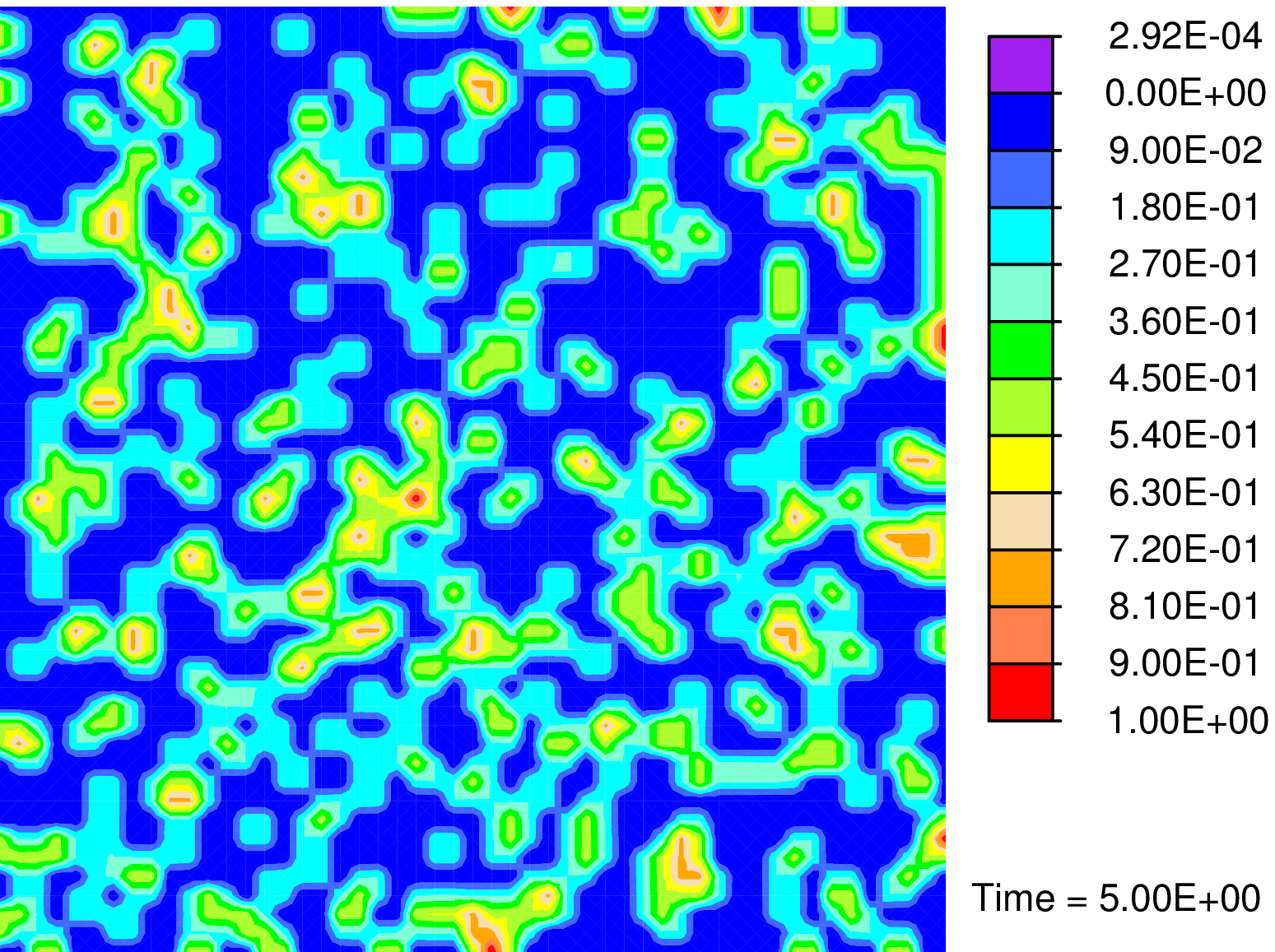} \hspace{1cm}
 (d)\,\includegraphics[width=0.39\textwidth]{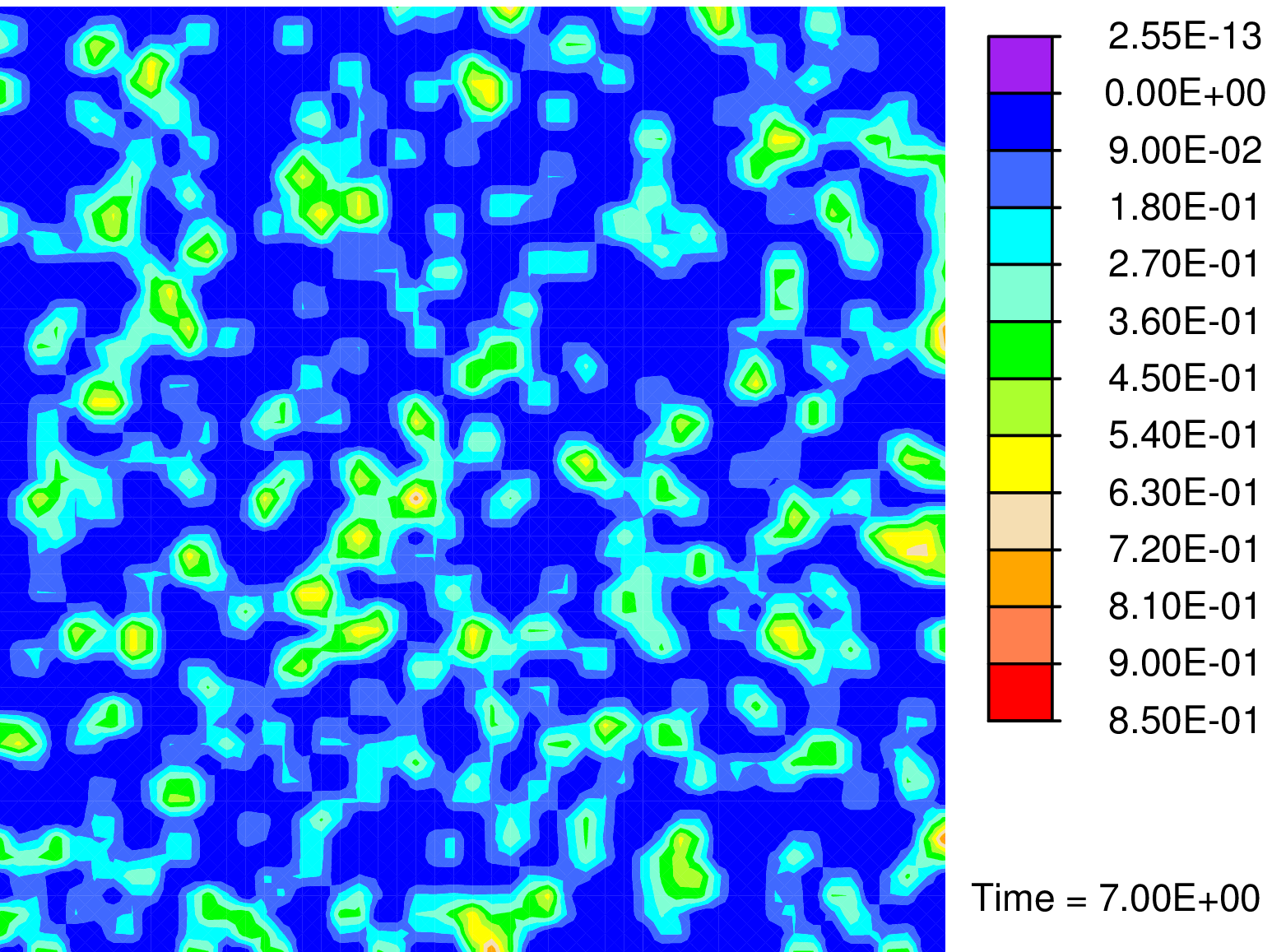}
 \caption{Cyclic tension test for a sample with a random initial microstructure. (a) Initial microstructure (type II). (b) Snapshot of the microstructure during the loading phase ($\bar{\lambda} = 4$). (c) State of the microstructure at the end of loading ($\bar{\lambda} = 6$). (d) Snapshot of the microstructure during the unloading phase ($\bar{\lambda} = 4$).}
 \label{test4}
\end{figure}

Selected numerical examples deal with the simulation of a tensile test performed on two-dimensional samples which depict the material microstructure. The elastic material parameters corresponding to rubber are chosen for the original amorphous structure  \cite{Shahzad2015,Maeda2015}, whereas the crystalline parameters are fitted to the experimental results by Toki et al. (2003) \cite{toki2003} and Candau et al. (2015) \cite{CANDAU2015244}. An overview of the material parameters is presented in Table \ref{mat_par}. In examples, the initial values of the network regularity are chosen to simulate specific cases of the material microstructure. Such an assumption is physically motivated, since areas with a higher regularity represent potential nuclei of crystal regions in real materials.
\begin{table}[h]
  \centering
  \renewcommand*{\arraystretch}{1.2}
  \begin{tabular}{llcc}
	\hline
	{\bf Elastic parameters}&&&\\
    \hline
    Bulk modulus & $K$ & $5\mathrm{E}8$ & Pa \\
    Shear modulus & $\mu$ & $4\mathrm{E}5$ & Pa \\
    Limiting network stretch & $\lambda_m$ & 2 & --\\
    \hline
	{\bf Crystalline parameters} &&\\
	\hline
	Coupling parameter & $k$ & $7\mathrm{E-}2$ & --\\
	Crystallization limit & $A$ & $1\mathrm{E}5$ & Pa\\
	Hardening parameter & $b_1$ &  $1.7\mathrm{E}5$ & Pa \\
	Softening parameter & $b_2$ &  $2\mathrm{E}5$ & Pa \\
	Parameter in function $f(\chi)$ & $\alpha$ &  $0.25$ & -- \\
	Parameter in function $f(\chi)$ & $\beta$ &  $0.5$ & -- \\
	\hline
  \end{tabular}
  \caption{Material parameters used in simulations.}
  \label{mat_par}	
\end{table}
\\
The set-up corresponding to the tensile test is shown in Fig. \ref{init} a. The chosen square sample has the dimensions $100 \times 100$  nm and is discretized by $50 \times 50$ quadrilateral elements.
The assumed sample is large enough to monitor and visualize the evolution of several crystals since their average size amounts to 10 nm \cite{mistry2014, doi:10.1021/ma5006843, huneau2011}.
The sample  thickness (1 nm) is significantly smaller than the remaining dimensions, which corresponds to a plane stress state problem. However, the application of the model to the 3D simulations is straightforward, since  the general 3D SIC-material model is proposed in previous sections. Vertical displacements prescribed at the horizontal boundaries linearly increase up to the maximal value of 250 nm, and thereafter linearly decrease to 0 (Fig. \ref{init} b). The displacement increment in both phases is set to $\left\vert \Delta \bar{u} \right\vert =$ 1E-2 nm. Here, the bar symbol indicates external influences. The prescribed stretch is calculated according to $\bar{\lambda} = (l+2\,\bar{u}) / l$. The total loading time amounts to 10 s and the time increment is $\Delta t =$ 2E-4 s.
\\ \\
The first two examples have an academic character and investigate the influence of different factors on the microstructure evolution if a simple initial configuration is assumed. The first case study focuses on the influence of the initial value of the regularity degree. Here, a tension test (Fig. \ref{init}) is performed on a sample with three dilute nuclei, each of them with another network regularity. The initial value of the regularity is set to \mbox{1E-4} at an element in the bottom part of the sample, to \mbox{1E-6} at an element in the middle of the sample and to \mbox{1E-8} at an element in the upper part of the sample. The numerical results for the complete loading cycle are shown in Fig. \ref{test1}. Figures \ref{test1} a-d show the gradual growth of crystalline regions during the loading stage. As expected, the crystalline regions build up and grow faster at the areas close to the element with a higher initial value of regularity. At the end of the loading phase, the full crystallization ($\chi\approx 1$) is achieved in the crystallites at the bottom part and in the middle of the sample. During the unloading stage (Figs. \ref{test1} e-f) the crystallinity degree gradually decreases. Eventually, crystalline regions disappear completely.
\\ \\
As a complement to the first case study, the second example monitors the interaction of the crystalline regions and the influence of the size of the nuclei (Fig. \ref{test2}). The initial values of the regularity at all nuclei are the same and amount to 1E-4. As a consequence, the network regularity grows equally fast and simultaneously reaches the maximum in all crystallites (Fig. \ref{test2} b). However, smaller crystalline regions vanish faster than the large ones (Fig. \ref{test2} c). This goes back to the contribution of the function $f(\chi)$ (Eq. \eqref{dB4}).
\\ \\
A further example simulates the tensile test for a sample with a randomly generated initial value of the network regularity (Fig. \ref{test3} a), which is a situation to be expected in a real polymer. The initial values are generated within range [0, 1E-2] and the tension test shown in Fig. \ref{init} is simulated. Three snapshots are chosen to illustrate the microstructure evolution: Fig. \ref{test3} b shows the microstructure corresponding to the external load $\bar{u} = 150$ nm ($\bar{\lambda} = 4$) during the loading phase, Fig. \ref{test3} c presents the situation at the end of the loading phase $\bar{u} = 250$ nm ($\bar{\lambda} = 6$) and  Fig. \ref{test3} d shows the microstructure for the external load $\bar{u} = 150$ nm ($\bar{\lambda} = 4$) during the unloading. The comparison of Figs. \ref{test3} b and \ref{test3} d shows that the crystalline regions are dominant in the second case although the same external load is applied. This clearly proves that the rate of the network regularity is higher during the loading stage than it is during the unloading phase. The color scale in Fig. \ref{test3} a is different from the color scale in Figs. \ref{test3} b-d which is necessary in order to visualize the initial microstructure.
\\ \\
An analogous cyclic test is performed for a sample with a modified initial random distribution as shown in Fig. \ref{test4} a. In both cases the same volume fraction of nuclei is chosen, their spatial distribution is however different. The evolution of crystalline regions shows a similar behavior as it does in the previous example; the growth of crystalline regions during the loading (Figs. \ref{test4} b and c) and the shrinkage during the unloading (Fig. \ref{test4} d).
\\ \\
Simulations performed on a sample with the random initial microstructure (Figs. \ref{test3} and \ref{test4}) provide a suitable basis for a comparison with the experimental results shown in Fig. \ref{fig1}. First, the focus is set on the investigation of $P_{22}$-component of the first Piola-Kirchhoff stress tensor and of its change versus the applied stretch $\bar{\lambda}$. Here, two aspects can be distinguished: the stress state at a single point and the effective stress state. 

\begin{figure}[h!]
  \psfrag{p2}[]{\footnotesize \, $P_{22}$, $P_{22}^{\mathrm{eff}}$ [MPa]}
  \psfrag{lam}[]{\footnotesize $\bar{\lambda}$}
  \psfrag{averaged}{\tiny $P_{22}^{\mathrm{eff}}$}
  \psfrag{point}{\tiny $P_{22}$}
  \psfrag{cd}[][][1][180]{\footnotesize Crystallinity degree [\%]}
  \psfrag{E}{\tiny Microstructure I}
  \psfrag{H}{\tiny Microstructure II}
  \centering
  \begin{minipage}{\linewidth}
  (a)\includegraphics[width=0.45\textwidth]{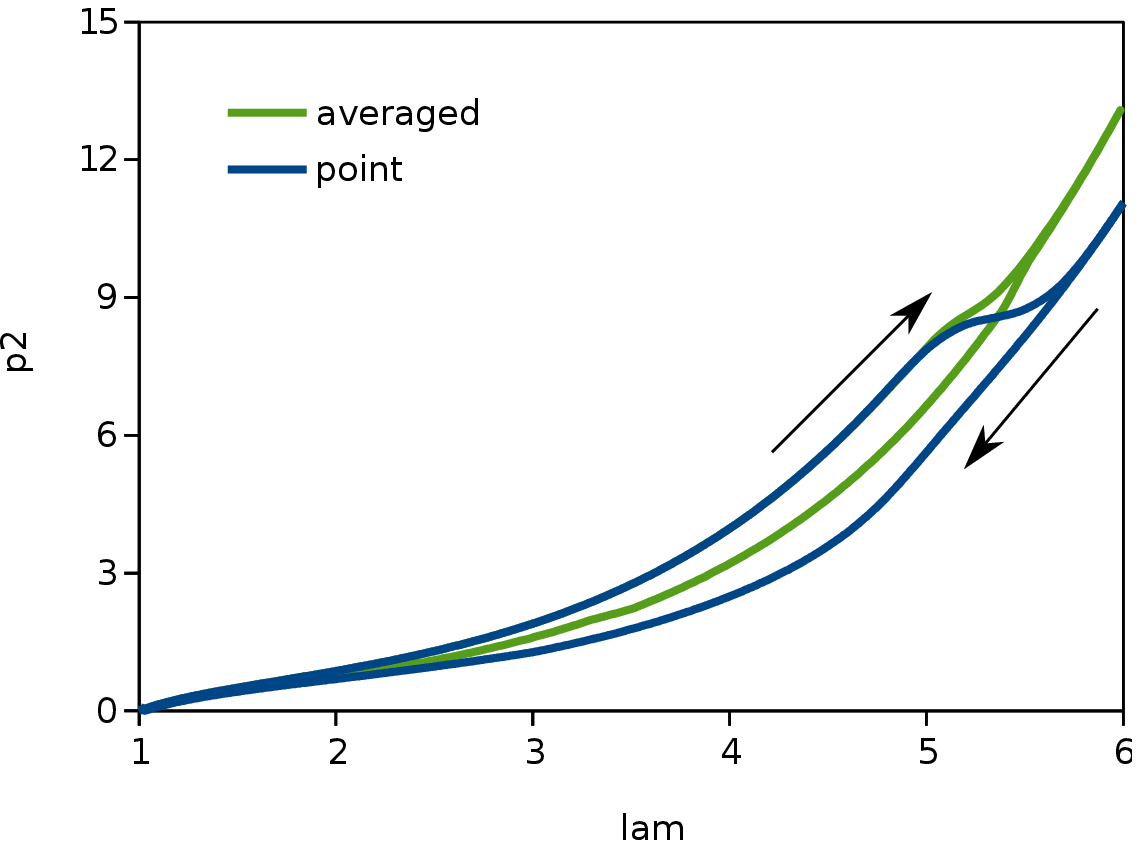}\hspace{1mm}
  (b)\raisebox{-0.5mm}{\includegraphics[width=0.45\textwidth]{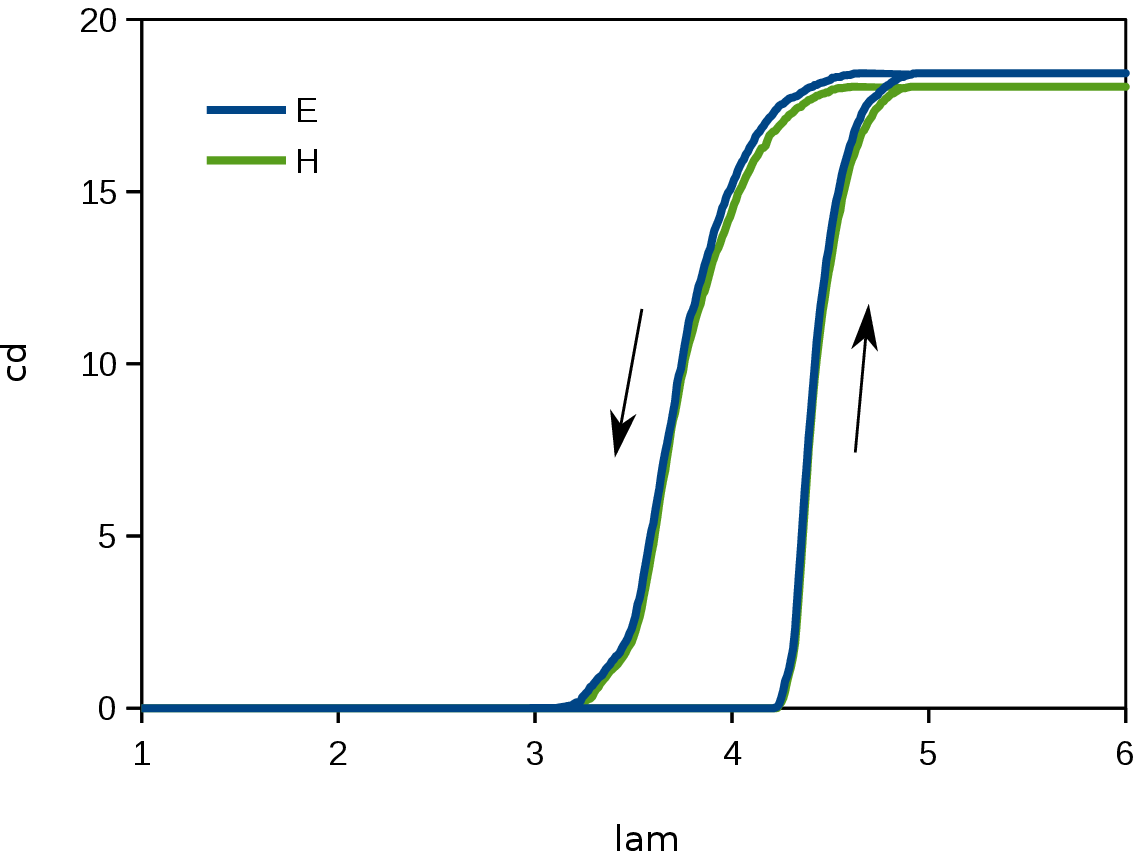}}
 \caption{(a) The $P_{22}$ component of the first Piola-Kirchhoff stress tensor at point P and its effective counterpart $P_{22}^{\mathrm{eff}}$. (b) The crystallinity degree versus the applied stretch for samples with different initial microstructures.}
  \label{average}
  \end{minipage}
\end{figure}

In order to display the stress state at a single point  that fully crystallizes, point P is chosen as presented in Fig. \ref{test3} a. The initial regularity at point P is higher than in the surrounding material which stipulates and accelerates the regularity evolution.
The stress at point P gradually increases up to the value $\bar{\lambda} = 5.1$ where stress growth
stagnates. Thereafter, the stress diagram builds a short plateau which finishes with a peak at $\bar{\lambda} = 5.6$. During the unloading phase, stress gradually decreases. The comparison of this diagram  (Fig. \ref{average} a, blue curve)  with the experimental results (Fig. \ref{fig1} a) shows an excellent agreement. In a further step, simulations shown in \ref{test3} are also used to evaluate the effective stresses $\bs{P}^{\mathrm{eff}}$ for the whole sample according to the principle of the volume averaging (Fig. \ref{average} a, green curve).  The hysteresis in this case becomes more narrow thus indicating that effective dissipation is smaller than  experimentally observed. This drawback is explained by the fact that the volume fraction of the amorphous phase is much larger than the volume fraction of the crystalline regions. However, the amorphous material behaves elastically and does not contribute to dissipation, which significantly decreases its effective value.
The same effect was observed in the work by Kroon (2010) \cite{kroon2010}, where the author introduces viscous effects in order to overcome this drawback. Other possible explanations are  effects of the interaction of the crystalline regions, effects of structural changes during the formation, such as the rotation of polymer chains, or contributions at the interface between the amorphous crystalline regions.
\\ \\
The final results deal with the change of the crystallinity degree (Fig. \ref{average} b). According to this diagram, the crystalline regions start to build at $\bar{\lambda} = 4.3$ and their volume fraction gradually  increases up to the value of $18 \%$ at the end of the loading phase. The crystallinity degree gradually decreases during the unloading stage and crystalline regions completely vanish at $\bar{\lambda} = 3.1$. The rate of change during the loading phase is higher than it is during the unloading stage. Both tension tests with different initial distributions of the regularity yield approximately the same results for the crystallinity degree. The archived numerical values show a excellent agreement with the experimental results (Fig. \ref{fig1} b). The diagram for a single point shown in Fig.  \ref{average} a can also be obtained by using a material point model. However, the advantage of the FEM simulations is that they give insight into the behavior of the whole sample and can be applied for the evaluation of effective quantities.

\section{Conclusion and outlook}
The present study focuses on the thermodynamically consistent mechanical modeling of the SIC phenomenon in unfilled polymers. The elastic behavior is described by the Arruda-Boyce model, whereas the evolution equation for the regularity of polymer chains and inelastic deformations due to the crystallization are derived by using the dissipation potential specifically proposed for this type of material behavior. These two internal variables are coupled by assuming a condition with the following implications. Firstly, the evolution of the regularity depends on the evolution direction defined in terms of deviatoric Mandel stresses. Secondly, the formulation of the free energy function enables the regularity to increase and decrease depending on the sign of the driving force rate. 
The SIC model proposed incorporates some elements typical of conventional plasticity with isotropic hardening. The internal variable $\chi$ describing regularity of chain alignment changes the material behavior of crystalline regions compared to the amorphous regions. The underlying idea is that the change of the regularity  contributes to the change of material behavior in the same manner as the accumulation of dislocations to the hardening. However, the similarities with the plasticity model are limited to the loading mode since the SIC model also simulates  microstructural changes during the unloading stage, which is not the case when classic plasticity is considered.
\\ \\
The application of the model has been illustrated  by several examples  dealing with monitoring the microstructure evolution  during a cyclic tension test. The initial configuration in examples has been varied for a better study of the influence of separate factors. The first two examples deal with a simple configuration and have an academic character. They investigate the influence of the initial value of the network regularity and of the interaction of  crystalline regions. The numerical results endorse the expectations that the higher network regularity leads to a faster development of the crystalline regions, as well as that the neighboring nuclei yield the merging of crystalline regions.\\ \\
The final examples simulate the behavior of samples corresponding to possible configurations of a real amorphous  polymer and enable the visualization of  growing and shrinkage  of crystalline regions during the loading and unloading stage respectively. The shrinkage of the regions is slower than their growth, as is experimentally observed.  In a post-processing step, the volume averaging procedure provides the diagrams depicting the effective polymer behavior. The diagram of crystallinity degree versus stretches shows an excellent agreement with the experimental results, whereas the stress-stretch diagram builds a hysteresis narrower than its experimental counterpart. This drawback indicates that an extension of the model is necessary and that additional aspects, possibly the interaction of crystalline regions, might contribute to the dissipative processes more significantly. \\ \\

Apart from the issues mentioned, the developed model also gives rise to some other investigations. In a first step, the proposals for the Helmholtz energy and the dissipation potential can be extended by considering further effects occurring in filled and unfilled rubbers. These can be the Mullins effect, a deformation state beyond the elastic limit, induced anisotropy and thermal influences. In addition, the model proposed can be coupled to the phase-field approach in order to represent  the two-phasic nature of material in a more realistic way. Internal variable $\chi$ would correspond to the order parameter in that case, and its evolution could be controlled by the same dissipation potential as proposed in the present work.  A combination with the phase-field method would certainly be an attractive topic for the future work, since this strategy has already found application in many research areas where the microstructure evolution plays an important role \cite{Takaki,doi:10.1063/1.4923226}.

\section*{Acknowledgment}
We gratefully acknowledge the financial support of the German Research Foundation (DFG), research grant KL 2678/7-1. We also thank Prof. J. Tiller and Dr. F. Katzenberg for helpful discussions.

\bibliography{references}

\appendix

{\section{FE framework for nonlinear materials at finite deformations}}
\label{appendix}
The SIC model presented in Sect. \ref{sec2} is implemented into an FE code by using the standard framework relying on the strong formulation of the boundary value problem \cite{wriggers2008nonlinear}
\begin{linenomath*}
\begin{equation}
  \mathrm{Div} \bs{P} + \rho \, \bs{b} = \bs{0} \text{ ,} \quad \bs{t} = \bs{P} \cdot \bs{n} \text{ on } \partial \mathcal{B}^{\bs{t}} \text{ ,} \quad \bs{u} = \bar{\bs{u}} \text{ on } \partial \mathcal{B}^{\bs{u}} \text{ .}
\end{equation}
\end{linenomath*}
Here, $\rho$ denotes the density, $\bs{b}$ is the body force, $\bs{t}$ is the traction, $\bs{n}$ is the surface normal and $\bs{u}$ is the displacement. If $\mathcal{B}$ is the body, the Neumann boundary conditions act on surface $\partial \mathcal{B}^{\bs{t}}$, whereas the Dirichlet boundary conditions act on surface $\partial \mathcal{B}^{\bs{u}}$ with prescribed displacement $\bar{\bs{u}}$. The transformation of the strong form into the weak form requires the following two steps: multiplication of the strong form by a test function $\delta \bs{u}$, commonly referred to as virtual displacements, and the integration over the body. Finally, the weak form of the balance of linear momentum is obtained by using integration by parts and the divergence theorem
\begin{linenomath*}
\begin{equation}
  \int_{\mathcal{B}} \nabla_{\bs{X}} \delta \bs{u} \colon \bs{P} \, \mathrm{d} V - \int_{\mathcal{B}} \delta \bs{u} \cdot \rho \, \bs{b} \, \mathrm{d}V - \int_{\partial \mathcal{B}^{\bs{t}}} \delta \bs{u} \cdot \bs{t} \, \mathrm{d} A = 0 \text{ .}
  \label{wf}
\end{equation}
\end{linenomath*}
In a next step, the body $\mathcal{B} \approx \bigcup_{\mathrm{e}=1}^{n_{\mathrm{el}}} \mathcal{B}^{\mathrm{e}}$ is spatially disrcretized into a finite number of elements $n_{\mathrm{el}}$ and integrals in Eq. \eqref{wf} are transformed into a sum of integrals over single elements $\mathcal{B}^{\mathrm{e}}$
\begin{linenomath*}
\begin{equation}
 \sum_{\mathrm{e} = 1}^{n_{\mathrm{el}}} \left\{ \int_{\mathcal{B}^{\mathrm{e}}} \nabla_{\bs{X}} \delta \bs{u}^{\mathrm{e}} \colon \bs{P} \, \mathrm{d} V - \int_{\mathcal{B}^{\mathrm{e}}} \delta \bs{u}^{\mathrm{e}} \cdot \rho \, \bs{b} \, \mathrm{d}V - \int_{\partial \mathcal{B}^{\mathrm{e} \, \bs{t}}} \delta \bs{u}^{\mathrm{e}} \cdot \bs{t} \, \mathrm{d} A \right\} = 0 \text{ .}
  \label{dwf}
\end{equation}
\end{linenomath*}
The last integral is only active in elements where the traction boundary conditions are prescribed. \\
\\
For further analysis, a nonlinear 2D-quadrilateral element with four nodes is selected. Here, the C$^0$-continuous shape functions $N^A$, of nodes $A=1,\dots,n_{\mathrm{en}}$ are used to map the physical and the parametric spaces. The approximation of test functions by elementwise polynomials is then written as 
\begin{linenomath*}
\begin{equation}
  \delta \bs{u}^{\mathrm{e}} = \sum_{A=1}^{n_{\mathrm{en}}} \delta \bs{u}^{\mathrm{e} \, A} \, N^A \quad \text{and} \quad \nabla_{\bs{X}} \delta \bs{u}^{\mathrm{e}} = \sum_{A=1}^{n_{\mathrm{en}}} \delta \bs{u}^{\mathrm{e} \, A} \otimes \nabla_{\bs{X}} N^A \text{ ,}
  \label{shape}
\end{equation}
\end{linenomath*}
where $\delta \bs{u}^{\mathrm{e} \, A}$ is the value of the virtual displacement at node $A$ of element e. The displacement field $\bs{u}^{\mathrm{e}}$ is approximated in the same way. The insertion of Eq. \eqref{shape} into Eq. \eqref{dwf} reads
\begin{linenomath*}
\begin{equation}
  \begin{split}
  & \sum_{\mathrm{e} = 1}^{n_{\mathrm{el}}} \left\{  \sum_{A=1}^{n_{\mathrm{en}}} \delta \bs{u}^{\mathrm{e} \, A} \cdot \left[ \bs{f}_{\mathrm{int}}^{\mathrm{e} \, A} - \bs{f}_{\mathrm{vol}}^{\mathrm{e} \, A} - \bs{f}_{\mathrm{sur}}^{\mathrm{e} \, A} \right] \right\} = 0 \text{ ,} \\
  & \bs{f}_{\mathrm{int}}^{\mathrm{e} \, A} = \int_{\mathcal{B}^{\mathrm{e}}} \bs{P} \cdot \nabla_{\bs{X}} N^A \, \mathrm{d} V
\text{ ,} \quad \bs{f}_{\mathrm{vol}}^{\mathrm{e} \, A} = \int_{\mathcal{B}^{\mathrm{e}}} \rho \, N^A \, \bs{b} \, \mathrm{d}V \text{ ,} \quad \bs{f}_{\mathrm{sur}}^{\mathrm{e} \, A} = \int_{\partial \mathcal{B}^{\mathrm{e} \, \bs{t}}} N^A \, \bs{t} \, \mathrm{d} A
  \end{split}
  \label{dwf_contr}
\end{equation}
\end{linenomath*}
with the contributions of the internal forces $\bs{f}_{\mathrm{int}}^{\mathrm{e} \, A}$, the volume forces $ \bs{f}_{\mathrm{vol}}^{\mathrm{e} \, A}$ and the surface tractions $\bs{f}_{\mathrm{sur}}^{\mathrm{e} \, A}$. Furthermore, the element contributions (Eq. \eqref{dwf_contr}) are assembled to a global system of equations under consideration of kinematic compatibility
\begin{linenomath*}
\begin{equation}
  \delta \bs{u}^T \cdot \bs{r} \left( \bs{u} \right) = 0 \text{ , } \quad \bs{r} \left( \bs{u} \right) = \bs{f}_{\mathrm{int}} \left( \bs{u} \right) - \bs{f}_{\mathrm{vol}} - \bs{f}_{\mathrm{sur}} \text{ .}
\label{resi}  
\end{equation}
\end{linenomath*}
Here, it is assumed that the external loads $ \bs{f}_{\mathrm{vol}}$ and $\bs{f}_{\mathrm{sur}}$ are independent of the deformation map, so-called dead loads. \\
\\
Equation \eqref{resi} defines a nonlinear system of equations which can be solved by using different techniques. The commonly used Newton-Raphson method, for example, linearizes the problem as follows 
\begin{linenomath*}
\begin{equation}
  \bs{J} \left( \bs{u}^k \right) \, \Delta \bs{u} = - \bs{r} \left( \bs{u}^k \right) \text{ ,} \quad \bs{u}^{k+1} = \bs{u}^k + \Delta \bs{u} \text{ .}
\end{equation}
\end{linenomath*}
In the previous expression, the Jacobian matrix $\bs{J}$ is calculated according to
\begin{linenomath*}
\begin{equation}
  \bs{J} = \frac{\partial \bs{r}}{\partial \bs{u}} = \frac{\partial  \bs{f}_{\mathrm{int}}}{\partial \bs{u}} \text{ ,}
  \label{jm}
\end{equation}
\end{linenomath*}
which also can be presented in the index notation
\begin{linenomath*}
\begin{equation}
  \frac{ \partial f_{\mathrm{int} \, i}^{\mathrm{e} \, A}}{\partial u_j} = \int_{\mathcal{B}^{\mathrm{e}}} \nabla_{X_k} N^A \,  \frac{\partial P_{i \, k}}{\partial F_{l \, m}} \frac{\partial F_{l \, m}}{\partial u_j} \, \mathrm{d} V \text{ ,} \quad F_{l \, m} = \sum_{B=1}^{n_{\mathrm{en}}} (X_l^{\mathrm{e} \, B} + u_l^{\mathrm{e} \, B}) \, \nabla_{X_m} N^B
\end{equation}
\end{linenomath*}
such that the tangent stiffness matrix $\bs{K}^{e \, A B}$ is written as
\begin{linenomath*}
\begin{equation}
  K^{e \, A B}_{i \, j} := \int_{\mathcal{B}^{\mathrm{e}}} \nabla_{X_k} N^A \,  C_{i \, k \, j \, m} \, \nabla_{X_m} N^B \, \mathrm{d} V  \text{ .}
  \label{tsm}
\end{equation}
\end{linenomath*}
Finally, the summation of the contributions Eq. \eqref{tsm} over $A$ and $B$ in the process of assembling the elements yields the global Jacobian matrix $\bs{J}$. Further details on the conventional assembly process can be found in the standard literature on FE analysis \cite{bathe2006finite,zie00}. \\
\\
The framework previously described corresponds to a purely elastic process without any dissipation, which is not the case if the SIC is simulated. Here, the decomposition \eqref{FF} and its influence on the definitions of stresses and tangent matrices has to be taken into consideration. For the particular case of SIC, the definitions of the first Piola-Kirchhoff stress tensor and material tensor turn into
\begin{linenomath*}
\begin{align}
    & \bs{P} = \frac{\partial \Psi}{\partial \bs{F}} = \frac{\partial \Psi^{\mathrm{e}}}{\partial \bs{F}^{\mathrm{e}}} \cdot {\bs{F}^{\mathrm{c}}}^{-T} = \frac{\partial \Psi^{\mathrm{e}}}{\partial \bs{C}^{\mathrm{e}}} \colon \frac{\partial \bs{C}^{\mathrm{e}}}{\partial \bs{F}^{\mathrm{e}}} \cdot {\bs{F}^{\mathrm{c}}}^{-T} = 2 \, \bs{F}^{\mathrm{e}} \cdot \frac{\partial \Psi^{\mathrm{e}}}{\partial \bs{C}^{\mathrm{e}}} \cdot {\bs{F}^{\mathrm{c}}}^{-T} \text{ ,} \\[2ex]
\begin{split}
  & C_{i \, k \, j \, m} = \frac{\partial P_{i \, k}}{\partial F_{j \, m}} = \frac{\partial P_{i \, k}}{\partial F^{\mathrm{e}}_{l \, n}} \frac{\partial F^{\mathrm{e}}_{l \, n}}{\partial F_{j \, m}} \\
  & \phantom{C_{i \, k \, j \, m}} = 2 \, \delta_{i \, j} \, {F^{\mathrm{c}}}^{-1}_{m \, l} \frac{\partial \Psi^{\mathrm{e}}}{\partial C^{\mathrm{e}}_{l \, n}} {F^{\mathrm{c}}}^{-T}_{n \, k} + 4 \, F^{\mathrm{e}}_{i \, l} \, {F^{\mathrm{c}}}^{-1}_{k \, n} \frac{\partial^2 \Psi^{\mathrm{e}}}{\partial C^{\mathrm{e}}_{l \, n} \, \partial C^{\mathrm{e}}_{o \, p}} F^{\mathrm{e}}_{j \, o} \, {F^{\mathrm{c}}}^{-1}_{m \, p} \text{ .}
\end{split}
\end{align}
\end{linenomath*}
Both quantities depend on the elastic energy (Eq. \eqref{ab}). Their evaluation requires the known deformation due to the crystallization $\bs{F}^{\mathrm{c}}$.

\end{document}